\documentclass[prd,onecolumn,notitlepage,showpacs,preprintnumbers,amsmath,amssymb,nofootinbib,APS,10pt,superscriptaddress]{revtex4-2}

\usepackage{dcolumn}
\usepackage{bm}
\usepackage{xcolor}
\usepackage[spanish,english]{babel}
\usepackage[utf8]{inputenc}
\usepackage{amsmath,amssymb,amsfonts,latexsym,cancel}
\usepackage{graphicx}
\usepackage{color}
\usepackage{soul}
\usepackage[normalem]{ulem}
\usepackage{hyperref}
\usepackage{amsmath}
\usepackage{slashed}
\usepackage{float}

\allowdisplaybreaks

\newcommand{\be}{\begin{equation}}
\newcommand{\ee}{\end{equation}}
\newcommand{\bea}{\begin{eqnarray}}
\newcommand{\eea}{\end{eqnarray}}

\begin{document}

\title{{\bf Equivalence of the adiabatic expansion and Hadamard renormalization \\ for a charged scalar field}}

\author{Silvia Pla}\email{silvia.pla\_garcia@kcl.ac.uk}

\affiliation{Department of Physics, King's College London, Strand Building, Strand Campus, Strand, London.  WC2R 2LS United Kingdom}

\affiliation{Departamento de Fisica Teorica and IFIC, Centro Mixto Universidad de Valencia-CSIC. Facultad de Fisica, Universidad de Valencia, Burjassot-46100, Valencia, Spain.}

\affiliation{Consortium for Fundamental Physics, School of Mathematics and Statistics,
Hicks Building, Hounsfield Road, Sheffield. S3 7RH United Kingdom}

\author{Elizabeth Winstanley}\email{E.Winstanley@sheffield.ac.uk}

\affiliation{Consortium for Fundamental Physics, School of Mathematics and Statistics,
Hicks Building, Hounsfield Road, Sheffield. S3 7RH United Kingdom}

\date{\today}

\begin{abstract}
We examine the relationship between three approaches (Hadamard, DeWitt-Schwinger and adiabatic) to the renormalization of expectation values of field operators acting on a charged quantum scalar field.
First, we demonstrate that the DeWitt-Schwinger representation of the Feynman Green's function is a particular case of the Hadamard representation.
Next, we restrict attention to a spatially flat Friedmann-Lema\^itre-Robertson-Walker universe with time-dependent, purely electric, background electromagnetic field, considering two, three and four-dimensional space-times. 
Working to the order required for the renormalization of the stress-energy tensor (SET), we find the adiabatic and DeWitt-Schwinger expansions of the Green's function when the space-time points are spatially separated. 
In two and four dimensions, the resulting DeWitt-Schwinger and adiabatic expansions are identical.
In three dimensions, the DeWitt-Schwinger expansion contains terms of adiabatic order four which are not necessary for the renormalization of the SET and hence absent in the adiabatic expansion. 
The equivalence of the DeWitt-Schwinger and adiabatic approaches to renormalization in the scenario considered is thereby demonstrated in even dimensions. In odd dimensions the situation is less clear and further investigation is required in order to determine whether adiabatic renormalization is a locally covariant renormalization prescription.
\end{abstract}

\maketitle

\section{Introduction}
\label{sec:intro}
In the absence of a full theory of quantum gravity, the study of quantum fields on a fixed, curved, space-time background has revealed many deep phenomena \cite{birrell-davies,fulling-book,wald,parker-toms}.
One of the most important of these is the production of quantum particles in an expanding universe \cite{Parker:1968mv,Parker:1969au,Parker:1971pt,Parker:2012at}, which leads to the subsequent generation of classical perturbations in the very early universe. 

For a specific quantum field, once a suitable quantum state has been specified, the properties of that state can be studied via the evaluation of expectation values of observables.
One of the key observables for any quantum field is the stress-energy tensor (SET) operator ${\hat {T}}_{\mu \nu }$, since
the expectation value of this quantity governs the back-reaction of the quantum field on the space-time geometry via the semiclassical Einstein equations.
The SET operator, in common with many observables, involves products of the quantum field operator evaluated at the same space-time point.
This means that a naive computation of its expectation value will give a divergent result.  
Therefore some kind of regularization (isolating the divergences) and renormalization (removing the divergences) scheme is required.

Of the many different renormalization prescriptions in the literature (see, for example, \cite{birrell-davies,fulling-book,parker-toms}), adiabatic renormalization is particularly well-adapted for finding expectation values on cosmological space-times. 
Assuming that the spatial geometry is flat, homogeneous and isotropic, a free classical field can be expanded into plane wave modes of fixed momentum, multiplied by functions of time. 
The short-distance singularity in the expectation values of operators thus corresponds to a high-momentum divergence in such expectation values when expressed as mode sums.

To renormalize the resulting divergence, an adiabatic expansion of the modes is performed, valid for large momenta. 
This adiabatic expansion depends on the background space-time geometry, but not on the quantum state of the field. 
Expectation values of operators are renormalized by subtracting from their mode sum expression sufficient terms in the adiabatic expansion of the modes to yield finite quantities. 
The adiabatic expansion of the modes can be found via an iterative method, with each order in the expansion given by algebraic expressions of increasing complexity, depending on the scale factor and its derivatives, the mode frequency and momentum, and quantities appearing in the classical scalar field equation such as the field mass.
One key advantage of the method is that the renormalization is performed mode-by-mode, with the resulting mode sums being of a suitable form for numerical computation.

Adiabatic renormalization was originally developed for  neutral scalar fields on cosmological space-times \cite{Fulling:1974zr,Parker:1974qw,Fulling:1974pu,Birrell:1978,Bunch:1978gb,Bunch:1980vc} and further improved in \cite{Anderson:1987yt} (see also~\cite{birrell-davies,fulling-book,parker-toms}). 
It has been extended spin-half fields \cite{Landete:2013lpa,Landete:2013axa,delRio:2014cha}, scalar and spin-half fields with a Yukawa coupling \cite{Yukawa:2017}, 
and scalar fields with a self-coupling \cite{Ferreiro:2022hik}.
Recently, time-dependent electric field backgrounds have also been incorporated into the adiabatic approach, for both scalar and Dirac fields
\cite{FN:2018,BarberoG:2018oqi,BNP:2020,Ferreiro:2018qzr}.

The adiabatic approach also serves to characterize the ultraviolet behaviour (in momentum space) of admissible quantum states through the so-called adiabatic condition (for subtle issues see \cite{Beltran-Palau:2022zvc}). However, one of its disadvantages is that it only applies, 
by construction, to homogeneous space-times \cite{Birrell:1978}. 
It is therefore useful to also have a more general framework for renormalization, valid for any space-time background and any quantum state.

One such framework is Hadamard renormalization. 
In this approach, expectation values are computed using the Green's function of the quantum field. 
For example, to find the expectation value of the SET for a free quantum field, a second order linear differential operator is applied to the Green's function. 
The Green's function depends on two space-time points, and is regular providing these two points are distinct. 
However, it is divergent in the coincidence limit. 
In Hadamard renormalization, the Hadamard representation of the Green's function is considered. 
The divergent terms present in the Hadamard representation of the Green's function are known as the Hadamard parametrix, which 
depends only on the background space-time geometry and the properties of the quantum field under consideration, and not on the quantum state. 
Renormalization of expectation values is achieved by subtracting the Hadamard parametrix from the Green's function, applying the appropriate differential operator (if necessary) and then bringing the space-time points together. 

For a general space-time background, the Hadamard parametrix is typically given as a covariant asymptotic series expansion in derivatives of the square of the geodesic distance between the two space-time points on which the Green's function depends. The coefficients in this expansion depend on the background geometry and parameters in the classical field equation.
Explicit expressions for the expansion coefficients have been given for  neutral \cite{Decanini:2005eg} and charged \cite{Balakumar:2019djw} scalar fields. 
The Hadamard prescription has also been developed for fermions \cite{Najmi:1984erj,Hollands:1999fc,Dappiaggi:2009xj,Lewis:2019iwu}, 
the electromagnetic field \cite{Brown:1986tj}, the St\"uckelberg
massive electromagnetic field \cite{Belokogne:2015etf}, $p$-forms \cite{Folacci:1990eb}, gauge bosons \cite{Frob:2017gez} and one-loop quantum gravity \cite{Allen:1987bn} (see also~\cite{Frob:2017gez}, where a general linear covariant gauge is employed).

A natural question then arises: how does Hadamard renormalization compare with adiabatic renormalization when applied to scenarios where adiabatic renormalization is applicable? 
If the answers yielded by these two methods are to be physically relevant, one requires them to give results which are equivalent up to the well-known renormalization ambiguities (for example, the renormalized SET is unique only up to the addition of a local, conserved, tensor \cite{wald,Wald:1977up}). 

For a neutral scalar field, the equivalence of adiabatic and Hadamard renormalization has been shown via two steps.
First, lengthy calculations \cite{Birrell:1978,delRio:2014:DWS} have demonstrated that adiabatic renormalization gives identical answers to DeWitt-Schwinger renormalization in two and four space-time dimensions.  
Second, the DeWitt-Schwinger representation of the Green's function is proven to be a special case of the Hadamard parametrix for a neutral scalar field \cite{Decanini:2005gt}.
Since DeWitt-Schwinger renormalization is effected by subtracting the DeWitt-Schwinger representation of the Green's function, the equivalence of Hadamard and adiabatic renormalization is thereby verified \cite{delRio:2014:DWS,Hack:2012qf,Ferreiro:2020uno,BDNN:2021}. 
A similar approach has also demonstrated the equivalence of DeWitt-Schwinger and adiabatic renormalization for a neutral Dirac field \cite{delRio:2014:DWS}.

Our purpose in this paper is to investigate the equivalence of adiabatic and Hadamard renormalization for a charged scalar field on a cosmological geometry, using the above two ingredients. 
The first step is contained in Sec.~\ref{sec:HdWS}. There we show, using a straightforward generalization of the work of Ref.~\cite{Decanini:2005gt}, that the DeWitt-Schwinger representation of the charged scalar Green's function on a general background metric and electromagnetic field \cite{Herman:1995hm,Herman:1998dz} is a particular case of the Hadamard parametrix \cite{Balakumar:2019djw}.
This result is valid for any number of space-time dimensions. 
Next, in Sec.~\ref{sec:adiabatic} we review the adiabatic formalism for a charged scalar field on a flat 
Friedmann-Lema\^itre-Robertson-Walker (FLRW) universe with a background, time-dependent, electric field. 
Restricting attention to two, three and four dimensions respectively, in Secs.~\ref{sec:two}--\ref{sec:four}, we then follow the method in \cite{delRio:2014:DWS} to  explore, via an explicit computation, whether the adiabatic Green's function, for space-like separated points, is equivalent to the DeWitt-Schwinger representation for our particular background metric and electromagnetic potential.
In even dimensions, we find that the adiabatic Green's function is indeed equivalent to the DeWitt-Schwinger representation, and hence adiabatic renormalization is a locally covariant renormalization scheme. However, for odd dimensions, we find terms in the DeWitt-Schwinger expansion which are of higher adiabatic order than those required for the renormalization of the SET. As a result, it is not clear whether adiabatic renormalization is a locally covariant renormalization scheme in odd dimensions and further investigation is required.
Our conclusions are presented in Sec.~\ref{sec:conc}. 
Two appendices include some lengthy algebraic expressions 
which arise in the four-dimensional case and details of the geodesic distance and Van-Vleck-Morette determinant for general point-splitting on our background space-time.

\section{Hadamard/DeWitt-Schwinger renormalization}
\label{sec:HdWS}
In this section we briefly review the Hadamard and DeWitt-Schwinger representations of the Feynman Green's function for a charged scalar field. We extend the result of Ref. \cite{Decanini:2005gt} to show that the DeWitt-Schwinger representation 
for a charged scalar field is a special case of the Hadamard representation. 

\subsection{Hadamard representation of the Feynman Green's function}
\label{sec:Hadamard}

We consider a massive charged scalar field $\Phi$ coupled to a classical electromagnetic background, propagating in an $N$-dimensional space-time, satisfying the Klein-Gordon equation
\be
(D_\mu D^{\mu} - m^2 - \xi R)\Phi=0\, , \label{KG}
\ee
where $D_\mu \Phi=(\nabla_\mu - \mathrm{i} q A_\mu)\Phi$, with $A_\mu$ the electromagnetic vector potential and $q$ the scalar field charge, $R$ is the Ricci scalar, $m$ the scalar field mass and $\xi$ is a dimensionless coupling constant. We use the metric signature $(-,+,+,....)$. 
The Feynman Green’s function of the scalar field is a biscalar function of the space-time points $x$ and $x^{\prime }$ and satisfies the (inhomogeneous) scalar field equation
\be \label{eq:FG-equation}
\left[D_{\mu} D^{\mu}-m^{2}-\xi R\right] G_{\mathrm{F}}(x, x^{\prime} )=-[-g(x)]^{-\frac{1}{2}} \delta^{N}(x-x^{\prime} )\, ,
\ee
where $g(x)$ is the determinant of the space-time metric. We assume that the space-time point $x^{\prime }$ lies in a normal neighbourhood of the point $x$, so that there is a unique geodesic connecting the two points. We also assume that the field is in a Hadamard state. This assumption determines the form of the Feynman propagator for closely separated space-time points. 

The Hadamard expansion of the Feynman Green's function depends on the geodesic interval $\sigma(x,x^{\prime })$ that is defined by the equation  
\be
2\sigma= g_{\mu \nu}\sigma^{;\mu} \sigma^{;\nu}\, .
\ee
The form of the Hadamard representation of the Green's function $G_{\mathrm {H}}(x,x^{\prime })$ depends on the number of space-time dimensions \cite{Decanini:2005gt,Balakumar:2019djw}:
\be\label{eq:Hadamard-all}
-\textrm{i}G_{\mathrm{H}}(x,x^{\prime }) = 
\begin{cases}
{\displaystyle{\frac{1}{4\pi } \left\{  
V(x,x^{\prime }) \ln \left[\frac{\sigma(x,x^{\prime })}{\ell^2} +\textrm{i}\varepsilon \right]
+ W(x,x^{\prime })
\right\} }}
& N=2 \,,
\vspace{0.3cm}
\\
{\displaystyle{\frac{(N/2-2)!}{2(2\pi )^{N/2}} \left\{ 
\frac{U(x,x^{\prime })}{[\sigma (x,x^{\prime }) + \textrm{i} \varepsilon ]^{N/2-1} } + V(x,x^{\prime }) \ln \left[\frac{\sigma (x,x^{\prime })}{\ell^2}+\textrm{i}\varepsilon\right]
+ W(x,x^{\prime })
\right\}  }}
& N>2 {\mbox { even}}\, ,
\vspace{0.3cm}
\\
{\displaystyle{\frac{\Gamma(N/2-1)}{2(2\pi )^{N/2}} \left\{  
\frac{U(x,x^{\prime })}{[\sigma (x,x^{\prime }) + \textrm{i} \varepsilon ]^{N/2-1} }
+ W(x,x^{\prime })
\right\}  }} 
& N {\mbox { odd}} \, ,
\end{cases}
\ee
where $\ell$ is an arbitrary renormalization length scale. 
The biscalars $U(x,x^{\prime })$ and $V(x,x^{\prime })$ are purely geometric, while $W(x,x^{\prime })$ may depend on the quantum state. 
For a charged scalar field, the functions $U(x,x^{\prime })$, $V(x,x^{\prime })$ and $W(x,x^{\prime })$ are complex sesquisymmetric biscalars that are regular when $x^{\prime }\to x$, and can be expanded in powers of $\sigma(x,x^{\prime })$ as 
\begin{subequations}
\label{eq:powersH}
\bea
\label{eq:powersU}
U(x,x^{\prime })&=&\sum_{n=0}^{h}U_n(x,x^{\prime })\sigma^n(x,x^{\prime })\, , \\
\label{eq:powersV}
V(x,x^{\prime })&=&\sum_{n=0}^{\infty}V_n(x,x^{\prime })\sigma^n(x,x^{\prime })\, ,\\
\label{eq:powersW}
W(x,x^{\prime })&=&\sum_{n=0}^{\infty}W_n(x,x^{\prime })\sigma^n(x,x^{\prime })\, ,
\eea
\end{subequations}
where   $h=N/2-2$ for $N>2$ even and $h\to \infty$ for $N$ odd. 
The expansions in (\ref{eq:powersH}) are to be understood as asymptotic expansions which are only convergent in analytic space-times (in more general space-times these expansions can be modified to give convergent quantities \cite{Hollands:2001fb}). The recurrence relations for the Hadamard coefficients $U_n(x,x^{\prime })$, $V_n(x,x^{\prime })$ and $W_n(x,x^{\prime })$ can be directly obtained from \eqref{eq:FG-equation}. For general background configurations it is not possible to find a closed form for these Hadamard coefficients. However, for the renormalization of expectation values, it is very convenient to perform a covariant asymptotic expansion of the coefficients, namely
  \begin{subequations}
  \label{eq:taylorUVW}
  \bea 
  \label{eq:taylorU} U_{n}(x, x^{\prime})=\sum_{j=0}^{\infty} U_{n j \alpha_{1} \ldots \alpha_{j}}(x) \sigma^{; \alpha_{1}}(x, x^{\prime}) \ldots \sigma^{; \alpha_{j}}(x, x^{\prime})\, ,\\
  \label{eq:taylorV}
  V_{n}(x, x^{\prime} )=\sum_{j=0}^{\infty} V_{n j \alpha_{1} \ldots \alpha_{j}}(x) \sigma^{; \alpha_{1}}(x, x^{\prime}) \ldots \sigma^{; \alpha_{j}}(x, x^{\prime} )\, ,\\
    \label{eq:taylorW}  W_{n}(x, x^{\prime} )=\sum_{j=0}^{\infty} W_{n j \alpha_{1} \ldots \alpha_{j}}(x) \sigma^{; \alpha_{1}}(x, x^{\prime} ) \ldots \sigma^{; \alpha_{j}}(x, x^{\prime} )\, .
  \eea  
  \end{subequations}
The coefficients of these asymptotic expansions are symmetric tensors of rank $(0,j)$ that only depend on the space-time point $x$. 

For completeness we give below the explicit recurrence relations for the geometric Hadamard coefficients $U_n(x,x^{\prime })$, $V_n(x,x^{\prime })$ and $W_n(x,x^{\prime })$. These recurrence relations involve the Van-Vleck Morette determinant $\Delta^{\frac{1}{2}}$. This is related to the D’Alembertian of the geodesic interval by
\be
\nabla_{\mu} \nabla^{\mu} \sigma=N-2 \Delta^{-\frac{1}{2}} \Delta_{; \mu}^{\frac{1}{2}} \sigma^{; \mu}\, .
\ee
From the recurrence relations given below it is possible to obtain the coefficients of the asymptotic expansions (\ref{eq:taylorU}, \ref{eq:taylorV}), as explicitly shown in \cite{Balakumar:2019djw}. 
For any $N$, the Hadamard coefficient $W_{0}(x,x^{\prime })$ is not determined by the recurrence relations; it may depend on the quantum state under consideration.
Once $W_{0}(x,x^{\prime })$ has been specified, the remaining Hadamard coefficients $W_{n}(x,x^{\prime })$, $n=1,2,\ldots $ are uniquely determined by the recurrence relations. 

The Hadamard parametrix is the Green's function of the form (\ref{eq:Hadamard-all}) with a choice of the biscalar $W(x,x^{\prime })$.
In Hadamard renormalization, $W(x,x^{\prime})$ is set to vanish in the Hadamard parametrix \cite{Hollands:2001fb,Hollands:2014eia}.
The Hadamard parametrix contains all the short-distance singularities in the Green's function. 
Renormalization is then effected by subtracting the chosen Hadamard parametrix from the Feynman Green's function for the quantum state under consideration. 
This process removes the short-distance singularities in the Feynman Green's function and the space-time points can then be brought together to give a finite limit. The renormalized Feynman Green's function is thus:
\begin{equation}
    G_{\rm {R}}(x,x^{\prime } ) = G_{\rm {F}}(x,x^{\prime }) - G_{\rm {H}}(x, x^{\prime }).
    \label{eq:renorm}
\end{equation}
The expectation value of the scalar condensate of a charged scalar field is given by the coincidence limit of the renormalized Feynman Green's function; the expectation value of the current follows from taking one derivative of $G_{\rm {R}}(x,x^{\prime })$ before bringing the points together, and the SET expectation value requires two derivatives to be applied to $G_{\rm {R}}(x,x^{\prime })$ prior to taking the coincidence limit. 
The number of terms in the Hadamard expansion (\ref{eq:Hadamard-all}, \ref{eq:taylorUVW}) which are required to be subtracted from the Green's function depends on the expectation value under consideration and the number of space-time dimensions. 
However, it is always possible to also subtract higher-order terms which vanish in the coincidence limit.

\subsubsection{Recurrence relations for $N=2$}
\begin{subequations}
 For $N=2$ the coefficients $V_n(x,x^{\prime })$ satisfy, for $n=0$,
\be
0= \left[\sigma^{; \mu} D_{\mu}-\Delta^{-\frac{1}{2}} \Delta_{; \mu}^{\frac{1}{2}} \sigma^{; \mu}\right] V_{0}\, ,
\ee
with boundary condition $V_0(x,x)=-1$. For $n=0,1,2,\ldots $ the recurrence relations are
\be \label{eq:rec-V-n2}
0= \left[D_{\mu} D^{\mu}-\left(m^{2}+\xi R\right)\right] V_{n} 
+2(n+1)\left[\sigma^{; \mu} D_{\mu}-\Delta^{-\frac{1}{2}} \Delta_{; \mu}^{\frac{1}{2}} \sigma^{; \mu}+(1+n)\right] V_{n+1}\,.
\ee
The Hadamard coefficients $W_{n}(x,x^{\prime })$ satisfy the recurrence relations
\bea
\label{eq:rec-W_n2}
0 =  \left[D_{\mu} D^{\mu}-\left(m^{2}+\xi R\right)\right] W_{n}
+2(n+1)\left[\sigma^{; \mu} D_{\mu}-\Delta^{-\frac{1}{2}} \Delta_{; \mu}^{\frac{1}{2}} \sigma^{; \mu}+(1+n)\right] W_{n+1}\nonumber \\ 
+2\left[\sigma^{; \mu} D_{\mu}-\Delta^{-\frac{1}{2}} \Delta_{; \mu}^{\frac{1}{2}} \sigma^{; \mu}+2(1+n)\right] V_{n+1} \, ,
\eea 
for $n=0,1, 2, \ldots $.  
In order to find the renormalized SET, the expansion of the Hadamard coefficient $V_{0}(x,x^{\prime })$ is required up to and including terms containing $\sigma ^{;\alpha _{1} }\sigma ^{;\alpha _{2}}$ and the leading-order term in $V_{1}(x,x^{\prime })$ is also necessary.

\end{subequations}

\subsubsection{Recurrence relations for $N>2$ even}
\begin{subequations}
For $N>2$ even, the coefficients $U_n(x,x)$ satisfy, for $n=0$,
\be
0= \left[\Delta^{-\frac{1}{2}} \Delta_{; \mu}^{\frac{1}{2}} \sigma^{; \mu}-\sigma^{; \mu} D_{\mu}\right] U_{0}\, , 
\ee
with boundary condition $U_0(x,x)=1$, and for $n=0,1,2,\ldots $
\be \label{eq:rec-U-neven}
0=  \left[D_{\mu} D^{\mu}-\left(m^{2}+\xi R\right)\right] U_{n}  -2(n+2-N/2)\left[\Delta^{-\frac{1}{2}} \Delta_{; \mu}^{\frac{1}{2}} \sigma^{; \mu}-\sigma^{; \mu} D_{\mu}-(n+1)\right] U_{n+1}\,  .
\ee
The coefficients $V_n(x,x^{\prime })$ obey (for $n\ge 0$)
\be \label{eq:rec-V-neven}
0= \left[D_{\mu} D^{\mu}-\left(m^{2}+\xi R\right)\right] V_{n} +2(n+1)\left[\sigma^{; \mu} D_{\mu}-\Delta^{-\frac{1}{2}} \Delta_{; \mu}^{\frac{1}{2}} \sigma^{; \mu}+(N/2+n)\right] V_{n+1}\, .
\ee
The coefficient $V_0(x,x^{\prime })$ is obtained from the following boundary condition 
\be
0= 2\left[\sigma^{; \mu} D_{\mu}-\Delta^{-\frac{1}{2}} \Delta_{; \mu}^{\frac{1}{2}} \sigma^{; \mu}+(N/2-1)\right] V_{0}+\left[D_{\mu} D^{\mu}-\left(m^{2}+\xi R\right)\right] U_{N/2-2} 
\, .
\ee
For $n=0,1,\ldots $, the recurrence relation satisfied by the $W_{n}(x,x^{\prime })$ is
\bea 
0= \left[D_{\mu} D^{\mu}-\left(m^{2}+\xi R\right)\right] W_{n} 
+2(n+1)\left[\sigma^{; \mu} D_{\mu}-\Delta^{-\frac{1}{2}} \Delta_{; \mu}^{\frac{1}{2}} \sigma^{; \mu}+(n+N/2)\right] W_{n+1} \nonumber \\
+2\left[\sigma^{; \mu} D_{\mu}-\Delta^{-\frac{1}{2}} \Delta_{; \mu}^{\frac{1}{2}} \sigma^{; \mu}+(2 n+1+N/2)\right] V_{n+1} \, .
\eea 
\end{subequations}
In this case, renormalization of the SET requires knowledge of the Hadamard coefficients $U_{0}(x,x^{\prime })$, \ldots ,$U_{N/2-2}(x,x^{\prime })$, with a covariant asymptotic series expansion of each $U_{p}(x,x^{\prime })$ up and including terms containing $\sigma ^{;\alpha _{1}}\ldots \sigma ^{;\alpha _{N-2p}}$.
In addition, the Hadamard coefficients $V_{0}(x,x^{\prime })$ and 
$V_{1}(x,x^{\prime })$ are necessary; with a covariant asymptotic  series expansion of $V_{0}(x,x^{\prime })$ up to and including terms containing $\sigma ^{;\alpha _{1}}\sigma ^{;\alpha _{2}}$ and the leading-order term in $V_{1}(x,x^{\prime })$.

\subsubsection{Recurrence relations for $N$ odd}
\begin{subequations}
Finally, for $N$ odd, the coefficient $U_0(x,x^{\prime })$ satisfies
\be 
0 = \left[\sigma^{; \mu} D_{\mu}-\Delta^{-\frac{1}{2}} \Delta_{; \mu}^{\frac{1}{2}} \sigma^{; \mu}\right] U_{0} \, ,
\ee
with boundary condition $U_0(x,x)=1$, while for $n=0,1,2,\ldots $ we have
\be \label{eq:rec-U-nodd}
0 = \left[D_{\mu} D^{\mu}-\left(m^{2}+\xi R\right)\right] U_{n}+(2 n+3-(N-1))\left[\sigma^{; \mu} D_{\mu}-\Delta^{-\frac{1}{2}} \Delta_{; \mu}^{\frac{1}{2}} \sigma^{; \mu}+(n+1)\right] U_{n+1} \, .
\ee
In this case the recurrence relations satisfied by the Hadamard coefficients $W_{n}(x,x^{\prime })$ are given by:
\bea
0 =(n+1)(2 n+N) W_{n+1}+2(n+1) W_{n+1 ; \mu} \sigma^{; \mu} 
-2(n+1) W_{n+1} \Delta^{-1 / 2} \Delta^{1 / 2}_{ ; \mu} \sigma^{; \mu} 
+\left(\square_{x}-m^{2}-\xi R\right) W_{n} \,
,
\eea
\end{subequations}
for $n=0,1,\ldots $.
The renormalized SET can be found if the Hadamard  coefficients $U_{0}(x,x^{\prime })$, \ldots , $U_{N/2-1/2}(x,x^{\prime })$ are known. The required asymptotic  series expansion of $U_{p}(x,x^{\prime })$ contains terms up to and including those involving $\sigma ^{;\alpha _{1}}\ldots \sigma ^{;\alpha _{N-2p}}$.

\subsection{DeWitt-Schwinger representation of the Feynman Green's function}
\label{sec:DWS}
The DeWitt-Schwinger representation of the Feynman propagator $G_{\mathrm {DS}}(x,x^{\prime })$ is given by
\be \label{eq:GDS}
G_{\mathrm{DS}}(x, x^{\prime} )=\textrm{i} \int_{0}^{+\infty} H(s ; x, x^{\prime} ) d s\, ,
\ee
where the kernel $H(s;x,x^{\prime })$ satisfies the equation
\be
\left(\textrm{i} \frac{\partial}{\partial s}+D_{\mu}D^{\mu}-m^{2}-\xi R\right) H(s ; x, x^{\prime} )=0 \quad \text { for } \quad s>0
\ee
with boundary condition $H(s ; x, x^{\prime}) \rightarrow (-g)^{-\frac{1}{2}}\delta^{N}(x- x^{\prime})$ as $s\to 0$. For $s\to 0$ and $x^{\prime }$ near $x$ the function $H(s ; x, x^{\prime})$ admits the following expansion\footnote{In some references, the proper-time expansion of the Feynman Green's function is defined with an overall extra factor $\Delta(x,x^{\prime })^{1/2}$. }
\be \label{eq:kernel-DWSexpansion}
H(s ; x, x^{\prime} )=\textrm{i}\,(4 \pi \textrm{i} s)^{-N/2} \exp 
\left\{
\frac{\textrm{i}}{2s}\left[\sigma(x, x^{\prime})+\textrm{i} \varepsilon\right]-\textrm{i} m^{2} s
\right\} \left[ \sum_{n=0}^{+\infty} \mathcal{A}_{n}(x, x^{\prime} )(\textrm{i} s)^{n} \right] \, .
\ee
The  DeWitt-Schwinger  coefficients $\mathcal{A}_n$ are complex sesquisymmetric biscalars that are regular for $x^{\prime }\to x$ and admit covariant  asymptotic expansions of the form
\be
\mathcal{A}_n(x,x^{\prime })=\sum_{j=0}^{\infty} \mathcal{A}_{nj  \alpha_{1} \ldots \alpha_{j}}(x) \sigma^{; \alpha_{1}}(x, x^{\prime} ) \ldots \sigma^{; \alpha_{j}}(x, x^{\prime} )\, .
\ee
The DeWitt-Schwinger coefficients are defined by the recurrence relation
\be
 \left[ (n+1)+ \sigma^{; \mu}D_\mu - \Delta^{-1 / 2} \Delta^{1 / 2}_{ ; \mu} \sigma^{; \mu} \right] \mathcal{A}_{n+1} 
=\left(D_\mu D^{\mu}-\xi R\right)\mathcal{A}_{n} \quad \text { for } n =0,1,\ldots  .
\ee
For $n=-1$ we can use the equation above by noting that $\mathcal{A}_{-1}=0$. Furthermore, for $n=0$ we have the boundary condition $A_0(x,x)=1$. 

Following \cite{Decanini:2005gt}, it is useful to define a new sequence of geometric coefficients, that we call the mass-dependent DeWitt-Schwinger coefficients, $\widetilde{\mathcal{A}}_n(m^2;x,x^{\prime })$, satisfying the recurrence relations
\be
\label{eq:Arecrel}
 \left[ (n+1)+ \sigma^{; \mu}D_\mu - \Delta^{-1 / 2} \Delta^{1 / 2}_{ ; \mu} \sigma^{; \mu} \right] \widetilde{\mathcal{A}}_{n+1} 
=\left(D_\mu D^{\mu}-m^2-\xi R\right)\widetilde{\mathcal{A}}_{n} \quad \text { for } n =0,1,\ldots .
\ee
Setting $n=-1$ in (\ref{eq:Arecrel}), we find that ${\widetilde {\mathcal {A}}}_{0} =  {\mathcal {A}}_{0}$ and does not depend on the scalar field mass $m$.
The mass-dependent coefficients correspond to an alternative expansion of $H(s;x,x^{\prime })$ where the exponential mass term is not included as in \eqref{eq:kernel-DWSexpansion}. From the expansion of $e^{-\textrm{i}m^2s}$, it is straightforward to see that the relation between the standard  $\mathcal{A}_n(x,x^{\prime})$ and the mass-dependent $\widetilde{ \mathcal{A}}_n(m^2;x,x^{\prime})$ coefficients is given by
\begin{equation}
\widetilde{\mathcal{A}}_{n}(m^{2} ; x, x^{\prime} )=\sum_{k=0}^{n} \frac{(-1)^{k}}{k !}\left(m^{2}\right)^{k} \mathcal{A}_{n-k}(x, x^{\prime} )\, .
\end{equation}
The coefficients $\mathcal A_n$ can also be directly obtained from the $\widetilde{\mathcal{A}}_n$ as follows
\be \label{eq:AnhatAn}
\mathcal{A}_n(x,x^{\prime })=\widetilde{\mathcal{A}}_n(m^2=0;x,x^{\prime })\, .
\ee
The advantage of the mass-dependent DeWitt-Schwinger coefficients is that they can be directly related to the geometric coefficients of the Hadamard expansion, as we shall see in the next subsection.
As with the Hadamard coefficients, the mass-dependent DeWitt-Schwinger coefficients can be given as covariant asymptotic  series expansions \cite{Decanini:2005gt}.

In the next subsection we shall show that the DeWitt-Schwinger representation of the Feynman Green's function has the Hadamard form (\ref{eq:Hadamard-all}) for a particular choice of the biscalar $W(x,x')$.  
The Green's function is renormalized by subtracting $G_{\rm {H}}(x,x')$ with this choice of $W(x,x')$ using (\ref{eq:renorm}). 
As can be seen below, the DeWitt-Schwinger representation is a large-mass expansion of the Feynman Green's function, and the number of terms in this expansion which need to be subtracted from the Feynman Green's function to give finite renormalized expectation values depends on the number of space-time dimensions $N$.

\subsection{Relation between the Hadamard and DeWitt-Schwinger representations}

In this part of the section, we give the explicit relation between the DeWitt-Schwinger and Hadamard coefficients in $N$ space-time dimensions. The relation between the geometric coefficients $U_n(x,x^{\prime })$ and $V_n(x,x^{\prime })$ and the mass-dependent DeWitt-Schwinger coefficients ${\widetilde{\mathcal{A}}}_n(m^2;x,x^{\prime })$ can be obtained by direct comparison of their recurrence relations. The correspondence between the  Hadamard coefficients $W_n(x,x^{\prime })$ and the DeWitt-Schwinger coefficients $\mathcal{A}_n(x,x^{\prime })$ is more complex. Its derivation requires lengthy intermediate steps that are detailed in the Appendix A. of Ref.~\cite{Decanini:2005gt} for a neutral scalar field and for $N>2$. 
The key point of the demonstration is to rewrite the DeWitt-Schwinger expansion of the Feynman Green's function as an asymptotic expansion in the geodesic distance. To this end, the authors of Ref.~\cite{Decanini:2005gt} assume that it is possible to exchange the integral and the sum in \eqref{eq:GDS} [see also \eqref{eq:kernel-DWSexpansion}], then perform the integration over $s$, make a short-distance expansion of the resulting expressions and group terms with the same characteristics to  match \eqref{eq:Hadamard-all}. The extension to a charged scalar field and $N=2$ is straightforward, so here we only give the final results.

As in \cite{Decanini:2005gt}, we find that the DeWitt-Schwinger expansion of the Feynman Green's function corresponds to the Hadamard parametrix with a particular choice of the Hadamard coefficient $W_{0}(x,x^{\prime })$, which is undetermined in the Hadamard formalism.
Once this coefficient is fixed, the remaining Hadamard coefficients $W_{n}(x,x^{\prime })$ are uniquely determined by the relevant recurrence relations. 
While in Hadamard renormalization the Hadamard parametrix (\ref{eq:Hadamard-all}) with the choice $W_{0}(x,x^{\prime })=0$ is subtracted from the Feynman Green's function, in DeWitt-Schwinger renormalization the DeWitt-Schwinger representation of the Green's function is subtracted. 

Below, for ease of reference, we give the expressions for the Hadamard coefficients $U_{n}(x,x^{\prime })$ and $V_{n}(x,x^{\prime })$ in terms of the mass-dependent DeWitt-Schwinger coefficients ${\widetilde {\mathcal{A}}}_{n}(m^{2};x,x^{\prime })$ and for the Hadamard coefficients $W_{n}(x,x^{\prime })$ in terms of the Hadamard coefficients $V_{n}(x,x^{\prime })$ and the DeWitt-Schwinger coefficients ${\mathcal{A}}_{n}(x,x^{\prime })$. 
For $N>2$, these expressions are identical to those in \cite{Decanini:2005gt}, and all the dependence on the background electromagnetic potential is contained in the DeWitt-Schwinger coefficients. 
For completeness, we also provide the corresponding expressions for the $N=2$ case, which is not considered in \cite{Decanini:2005gt}. 
For all $N$, the Hadamard coefficients $W_{n}(x,x^{\prime })$ are divergent in the limit $m^{2}\to 0$, due to the singularity in this limit of the DeWitt-Schwinger representation of the Feynman Green's function.

\label{sec:H=DWS}

\subsubsection{$N=2$}
\begin{subequations}
Comparing  \eqref{eq:rec-V-n2} with \eqref{eq:Arecrel}, and taking into account their boundary conditions, the relation between the DeWitt-Schwinger coefficients $\widetilde {\mathcal{A}}_n$ and the Hadamard coefficients $V_n(x,x^{\prime })$ reads
\be
\label{eq:VnAn-N2}V_{n}(x, x^{\prime} )=\frac{(-1)^{n+1}}{2^{n}\, n !} \widetilde{\mathcal A}_{n}(m^{2} ; x, x^{\prime} ) 
\quad\text { for } n =0, 1, \ldots .
\ee
Furthermore, the $W_n(x,x^{\prime })$ Hadamard coefficients associated with the DeWitt-Schwinger representation are 
\begin{multline}
W_{n}(x, x^{\prime} )=\left[ \ln \left(\frac{m^{2}\ell ^{2}}{2}\right) -2\psi(n+1) \right] V_{n}(x, x^{\prime} ) 
\\ -\frac{(-1)^{n}}{2^{n} n !}\bigg[\sum_{k=0}^{n-1} \frac{(-1)^{k}\left(m^{2}\right)^{k}}{k !}\left(\sum_{p=k+1}^{n} \frac{1}{p}\right) \mathcal A_{n-k}(x, x^{\prime} )-\sum_{k=0}^{+\infty} \frac{k !}{\left(m^{2}\right)^{k+1}} \mathcal A_{n+1+k}(x, x^{\prime} )\bigg] ,
 \label{eq:Wn-N2}
\end{multline}
where $\ell ^{2}$ is a renormalization length scale introduced so that the argument of the logarithm is dimensionless and $\psi (n+1)$ is the digamma function.
In particular,  the DeWitt-Schwinger representation of the Feynman Green's function takes the Hadamard form with the following choice of the coefficient $W_{0}(x,x^{\prime })$:
\bea
W_{0}(x, x^{\prime} )=\left[ \ln \left(\frac{m^{2}\ell ^{2}}{2}\right) +2\gamma  \right] V_{0}(x, x^{\prime} ) 
+\sum_{k=0}^{+\infty} \frac{k !}{\left(m^{2}\right)^{k+1}} \mathcal A_{k+1}(x, x^{\prime} ) ,
\eea
\end{subequations}
where $\gamma $ is the Euler-Mascheroni constant. 
For $N=2$, evaluation of the renormalized SET requires the Hadamard coefficients $V_{0}(x,x^{\prime })$ and $V_{1}(x,x^{\prime })$ and hence knowledge of the DeWitt-Schwinger coefficients ${\mathcal {A}}_{0}(x,x^{\prime })$ and ${\mathcal {A}}_{1}(x,x^{\prime })$. 
These are the only terms retained in $W_{n}(x,x^{\prime })$ given above. This  corresponds to an expansion of $W_{0}(x,x^{\prime})$ up to and including terms of order $m^{-2}$.
While both $W_{n}(x,x^{\prime })$ \eqref{eq:Wn-N2} and the Hadamard parametrix \eqref{eq:Hadamard-all} depend on the arbitrary renormalization length scale $\ell $, when the form of $W(x,x^{\prime })$ in the DeWitt-Schwinger representation is substituted into the Hadamard form (\ref{eq:Hadamard-all}), the terms dependent on $\ell ^{2}$ cancel, so that the DeWitt-Schwinger representation of the Green's function is independent of $\ell $. 
The renormalization length scale $\ell $ has effectively been replaced by the scalar field mass $m$, which is possible only in the massive case.

\subsubsection{$N>2$ even}
\begin{subequations}
\label{eq:DWSn-even}
As before, the Hadamard coefficients $U_n(x,x^{\prime })$, $V_n(x,x^{\prime })$ for $N>2$ even can be directly obtained from the mass-dependent DeWitt-Schwinger coefficients by direct comparison between their recurrence relations 
[see Eqs.~\eqref{eq:rec-U-neven}, \eqref{eq:rec-V-neven} and \eqref{eq:Arecrel}]. This time we obtain 
\bea \label{eq:UnAn-even}
U_{n}(x, x^{\prime} )&=& \frac{(N / 2-2-n) !}{2^{n}(N / 2-2) !}  \widetilde{ \mathcal A}_n(m^2;x,x^{\prime })
\quad \text { for } n=0,1, \ldots, N / 2-2\,,  \\
\nonumber\\
\label{eq:VnAn-even}V_{n}(x, x^{\prime} )&=&\frac{(-1)^{n+1}}{2^{n+N / 2-1} n !(N / 2-2) !} \widetilde{\mathcal A}_{n+N / 2-1}(m^{2} ; x, x^{\prime} ) 
\quad\text { for } n =0,1,\ldots  .
\eea
The $W_n(x,x^{\prime })$ Hadamard coefficients in the DeWitt-Schwinger representation are given by  
\begin{multline} 
W_{n}(x, x^{\prime} )= \left[ \ln \left(\frac{m^{2}\ell ^{2}}{2}\right) -\{ \psi(n+1)+\psi(n+N/2 )\}  \right] V_{n}(x, x^{\prime} ) \\
-\frac{(-1)^{n}}{2^{n+N/ 2-1} n !(N/ 2-2) !}\bigg[\sum_{k=0}^{n+N/ 2-2} \frac{(-1)^{k}\left(m^{2}\right)^{k}}{k !}\left(\sum_{p=k+1}^{n+N/ 2-1} \frac{1}{p}\right) \mathcal A_{n+N/ 2-1-k}(x, x^{\prime} )
 \\
 -\sum_{k=0}^{+\infty} \frac{k !}{\left(m^{2}\right)^{k+1}} \mathcal A_{n+N/ 2+k}(x, x^{\prime} )\bigg] ,
\label{eq:Wn}
\end{multline}
so that the DeWitt-Schwinger representation of the Feynman Green's function is of the Hadamard form with the choice
\begin{multline} 
W_{0}(x, x^{\prime} )= \left[ \ln \left(\frac{m^{2}\ell ^{2}}{2}\right) +\gamma - \psi(N/2 ) \right] V_{n}(x, x^{\prime} )  \\
 -\frac{1}{2^{N/ 2-1} (N/ 2-2) !}\bigg[\sum_{k=0}^{N/ 2-2} \frac{(-1)^{k}\left(m^{2}\right)^{k}}{k !}\left(\sum_{p=k+1}^{N/ 2-1} \frac{1}{p}\right) \mathcal A_{N/ 2-1-k}(x, x^{\prime} )
-\sum_{k=0}^{+\infty} \frac{k !}{\left(m^{2}\right)^{k+1}} \mathcal A_{N/ 2+k}(x, x^{\prime} )\bigg] .
\end{multline}
\end{subequations}
In this case, to find the renormalized SET, the DeWitt-Schwinger coefficients ${\mathcal {A}}_{0}(x,x^{\prime })$, \ldots , ${\mathcal {A}}_{N/2}(x,x^{\prime })$ are required. This means that we keep terms up to and including those of order $m^{-2}$ in $W_{0}(x,x^{\prime })$. 
As in two space-time dimensions, we have introduced a renormalization length scale in the argument of the logarithm in \eqref{eq:Wn}, however, when \eqref{eq:Wn} is substituted into the Hadamard form \eqref{eq:Hadamard-all}, the resulting DeWitt-Schwinger Green's function does not depend on $\ell $.

\subsubsection{Relation for $N$ odd}
\begin{subequations}
Finally, for $N$ odd we find, by comparing Eqs. \eqref{eq:rec-U-nodd} and \eqref{eq:Arecrel}, 
\begin{equation}
U_{n}(x, x^{\prime} )=\frac{\Gamma(N / 2-1-n)}{2^{n} \Gamma(N / 2-1)} \widetilde{ \mathcal A}_n(m^2; x, x^{\prime })
\quad \text { for } n =0,1,\ldots .
\label{eq:UAtildeodd}
\end{equation}
In this case the $W_n(x,x^{\prime })$ Hadamard coefficients for the DeWitt-Schwinger representation take a slightly simpler form compared with $N$ even, namely:
\begin{multline}
\label{eq:Wn-DWS-odd}
W_{n}(x, x^{\prime} )=-\frac{(-1)^{n}}{2^{n+N / 2-1} n ! \Gamma(N / 2-1)}\left[\sum_{k=0}^{n+N / 2-3 / 2} \frac{(-1)^{k}\left(m^{2}\right)^{k+1 / 2}}{\Gamma(k+3 / 2)} \pi \mathcal A_{n+N / 2-3 / 2-k}(x, x^{\prime} )\right. \\
\left.-\sum_{k=0}^{+\infty} \frac{\Gamma(k+1 / 2)}{\left(m^{2}\right)^{k+1 / 2}} \mathcal A_{n+N / 2-1 / 2+k}(x, x^{\prime} )\right] .
\end{multline}
As for other values of $N$, the DeWitt-Schwinger representation of the Feynman Green's function is a particular case of the Hadamard representation, with the following choice of the Hadamard coefficient $W_{0}(x,x^{\prime })$:
\begin{multline}
W_{0}(x, x^{\prime} )=-\frac{1}{2^{N / 2-1} \Gamma(N / 2-1)}\left[\sum_{k=0}^{N / 2-3 / 2} \frac{(-1)^{k}\left(m^{2}\right)^{k+1 / 2}}{\Gamma(k+3 / 2)} \pi \mathcal A_{N / 2-3 / 2-k}(x, x^{\prime} )\right. \\
\left.-\sum_{k=0}^{+\infty} \frac{\Gamma(k+1 / 2)}{\left(m^{2}\right)^{k+1 / 2}} \mathcal A_{N / 2-1 / 2+k}(x, x^{\prime} )\right] .
\end{multline}
\end{subequations}
Finding the renormalized SET for $N$ odd involves the DeWitt-Schwinger coefficients ${\mathcal {A}}_{0}(x,x^{\prime })$, \ldots , ${\mathcal {A}}_{N/2-1/2}(x,x^{\prime })$, and thus an expansion of $W_{0}(x,x^{\prime })$ up to and including terms of order $m^{-1}$.

\section{Adiabatic renormalization}
\label{sec:adiabatic}
In this section we examine the adiabatic expansion of the Feynman Green's function for charged scalar fields in homogeneous and time-dependent backgrounds.
To this end, we consider again the charged scalar field $\Phi$ of the previous section, coupled to a classical, time-dependent electric background, whose vector potential is of the form  $A_\mu=(0,A(t),0,...,0)$, propagating in a $N$-dimensional flat FLRW universe with line element 
\be \label{FLRWmetric} ds^2=-dt^2+a(t)^2 d{\bf x}^2\, .\ee  
Due to spatial homogeneity, we can perform a mode expansion of the scalar field as follows 
    \be
               \Phi(t,{\bf x})=\int \frac{d^{(N-1)}k}{\sqrt{2(2\pi)^{N-1}a(t)^{N-1}}}\Big(b_{{\bf k}}e^{\mathrm{i}{\bf k}{\bf {\cdot}}{\bf x}}h_{{\bf k}}(t)+d^\dagger_{{\bf k}}e^{-\mathrm{i}{\bf k}{\bf {\cdot }}{\bf x}}h^*_{-{\bf k}}(t)\Big)\, .\label{h-mode-expansion}
               \ee
In \eqref{h-mode-expansion}, $b_{{\bf k}}$, $b_{{\bf k}}^\dagger$, $d_{{\bf k}}$ and $d_{{\bf k}}^\dagger$ are  the usual annihilation and creation operators satisfying the commutation relations $[b_{{\bf k}},b^\dagger_{{\bf k}'}]=\delta({\bf k}-{\bf k}')=[d_{{\bf k}},d^\dagger_{{\bf k}'}]$. %, 
The mode function $h_{{\bf  k}}(t)$ satisfies a second order ordinary differential equation, obtained from the Klein-Gordon equation \eqref{KG}
\begin{equation}
\ddot h_{{\bf  k}} +\left[\frac{k^2}{a^2} + m^2 -\frac{2 q k_{1} A(t)}{a^2} +\frac{q^2 A(t)^2}{a^2}+\chi(t)\right]h_{{\bf k}}=0\, ,
\label{motion-h}
\end{equation}
where $k^2=|{\bf k}|^2$, $k_1$ is the first component of the momentum ${\bf  k}$, 
               and                
               \be
              \chi(t)= (N-3)(1-N)\frac{\dot a^2}{4a^2} +(1-N)\frac{\ddot a}{2a}+\xi R \, ,
              \label{eq:chi}
               \ee
               together with the Wronskian condition 
               \be \label{Wronskian}
         \dot{h}_{{\bf  k}}^*h_{{\bf  k}}-h_{{\bf  k}}^*\dot{h}_{{\bf  k}}=2{\mathrm {i}}\, .
               \ee    
 From the mode expansion \eqref{h-mode-expansion} we can easily compute the vacuum expectation values of relevant operators in terms of the mode functions. For example, the formal vacuum expectation value of the two-point function (at coincidence) $\langle \Phi \Phi^\dagger \rangle$ is given by 
                \be
               \langle \Phi \Phi^\dagger\rangle =  \frac{1}{2 (2\pi)^{N-1} a(t)^{N-1}}\int d^{(N-1)}k \, |h_{{\bf  k}}|^2 \label{phi2}.
                \ee 
This quantity is manifestly ultraviolet (UV) divergent, and has to be renormalized in a consistent way, compatible with general covariance. For time-dependent backgrounds there is a very efficient and direct technique, that takes advantage of the isometries of the space-time: the adiabatic regularization method. It is based on the adiabatic expansion of the field modes, and works as follows.
                
Given a mode function $h_{{\bf k}}(t)$ it is possible to perform an adiabatic expansion in terms of the time-dependent background fields. For scalars it is based on the Wentzel-Kramers-Brillouin (WKB) expansion, 
\be \label{eq:WKBansatz}
h_{{\bf k}}\sim \frac{1}{\sqrt{\Omega_{{\bf k}}(t)}}e^{-\mathrm{i} \int_{t_0}^t  \Omega_{{\bf k}}(u) \, du},
\ee
where $ \Omega_{{\bf k}}$ can be expanded adiabatically in powers of the derivatives of $a(t)$, the function $A(t)$ and its derivatives as follows
\be
\Omega_{{\bf k}}=\sum_{n=0}^\infty\omega_{{\bf k}}^{(n)}.
\label{eq:Omega}
\ee
The adiabatic expansion is uniquely determined once we fix its leading order. We require this to be 
\be
\omega_{{\bf k}}^{(0)}\equiv \omega=\sqrt{k^2/a^2+m^2}. 
\ee
Therefore, we are implicitly assuming that the function $A(t)$ is a function of  adiabatic order one~\cite{Ferreiro:2018qzr}. Subsequent terms are obtained, by iteration, from the relation
\be
 \Omega_{{\bf k}}^2= \omega^2+\chi +\frac{q^2 A^2}{a^2} -\frac{2 q k_{1} A}{a^2} + \frac{3}{4}\frac{\dot  \Omega_{{\bf k}}^2}{ \Omega^2_{{\bf k}}}-\frac{1}{2}\frac{\ddot  \Omega_{{\bf k}}}{ \Omega_{{\bf k}}}\, ,
 \label{eq:Omegarel}
\ee
derived from the ansatz \eqref{eq:WKBansatz} and the mode equation \eqref{motion-h}. We recall here that $\chi(t)$ is a function of adiabatic order two (it contains two derivatives of $a$). Inserting the adiabatic expansion \eqref{eq:Omega} in \eqref{eq:Omegarel}, and grouping terms with the same adiabatic order, it is possible to obtain the $n$th coefficient from the lower order ones once the leading order term is defined. For example, the next two terms are
\be
\omega^{(1)}_{{\bf k}}=-\frac{k_{1} q A}{a^2 \omega}\,, \qquad
\omega^{(2)}_{{\bf k}}=\frac{q^2 A^2}{2 a^2 \omega} -\frac{\big(\omega^{(1)}_{\bf{k}}\big)^2}{2\omega}  + \frac{4 \chi \omega^2+3 (\dot \omega)^2-2 \omega \ddot \omega}{8\omega^3}\,.
\ee
 
The main advantage of the adiabatic expansion is that it captures, in its leading terms, the expected UV divergences of the vacuum expectation values of physical observables 
 and therefore, it can be used for renormalization.  
 The renormalized version of the two-point function 
 can be obtained by subtracting terms up to and including the ($N-2$)th-order of its adiabatic expansion, namely
\be
\langle \Phi \Phi^\dagger\rangle_{\textrm{ren}}=\frac{1}{2(2\pi)^{N-1}a(t)^{N-1}}\int d^{N-1}k\Big(|h_{{\bf k}}|^2-\sum_{j=0}^{N-2}(\Omega_{{\bf k}}^{-1})^{(j)}\Big).
\ee
The number of subtractions required depends on the space-time dimension $N$ and also on the scaling dimension of the operator. To renormalize the charge current we require terms up to $(N-1)$th adiabatic order, while renormalizing the SET involves subtracting up to the $N$th adiabatic order.

The first few orders of the adiabatic expansion of $|h_{\bf k}|^2$ are
\begin{subequations}
\bea
(\Omega^{-1}_{{\bf k}})^{(0)}&=&\frac{1}{\omega}\, ;\\
                (\Omega^{-1}_{{\bf k}})^{(1)}&=&\frac{k_1qA}{a^2 \omega^3} \, ;\\
                (\Omega^{-1}_{{\bf k}})^{(2)}&=&-\frac{5 m^4 \dot{a}^2}{8 a^2 \omega ^7}+\frac{m^2 \dot{a}^2}{2
   a^2 \omega ^5}+\frac{\dot{a}^2}{8 a^2 \omega ^3}+\frac{m^2
   \ddot{a}}{4 a \omega ^5}-\frac{\ddot{a}}{4 a \omega
   ^3}-\frac{\chi}{2 \omega ^3}+\frac{3k_1^2 q^2 A^2}{2a^4\omega^5}-\frac{ q^2 A^2}{2a^2\omega^3}\, ;\\
                 (\Omega^{-1}_{{\bf k}})^{(3)}&=&-\frac{k_1\,q \ddot{A}}{4 a^2 \omega^5}+\frac{k_1}{a}\frac{q\dot{A}}{a}  \left(\frac{5m^2 \dot{a}}{4 a \omega^7}-\frac{ \dot{a}}{4 a^3 \omega ^5}\right)+ \frac{q^3A^3}{a^3} \left(-\frac{3 k_1}{2 a \omega ^5}+\frac{5 k_1^3}{2 a^3 \omega ^7}\right) \nonumber \\
                &&+\frac{k_1}{a}\frac{qA}{a} \left(-\frac{35 m^4 \dot{a}^2}{8 a^2 \omega ^9}+\frac{5 m^2 \dot{a}^2}{2
   a^2 \omega ^7}+\frac{3 \dot{a}^2}{8 a^2 \omega ^5}+\frac{5 m^2
   \ddot{a}}{4 a \omega ^7}-\frac{3 \ddot{a}}{4 a \omega
   ^5}-\frac{3 \chi}{2  \omega ^5}\right)\,.
\eea
\end{subequations}
The lengthy explicit expression for $(\Omega^{-1}_{{\bf k}})^{(4)}$ is given in Appendix \ref{ap:detailsD4}.

 The mode expansion of the field also allows us to obtain the formal value of the two-point function at two spatially separated points with spatial parts ${\bf x}$ and ${\bf x} '$ (we define $\Delta {\bf x} = {\bf x} - {\bf x}'$). In particular, it is possible to compute the Feynman Green's function at time coincidence 
                    \bea
              -\mathrm{i} G_{\textrm{F}}(t,{\bf x}\, ; t,{\bf x}')=\langle \Phi(t,{\bf x}) \Phi^\dagger(t,{\bf x}')\rangle=\frac{1}{2 (2\pi)^{N-1} a^{N-1}}\int d^{(N-1)}k \, e^{\mathrm{i} {\bf k} {\bf {\cdot }}\Delta {\bf x} } |h_{{\bf k}}|^2 .\label{green-h}
                \eea
                Therefore, its adiabatic expansion to adiabatic order $n$ reads
                \bea
              -\mathrm{i}  G_{\textrm{Ad}}^{(n)}(t,{\bf x}\, ; t,{\bf x}')=\frac{1}{2 (2\pi)^{N-1} a^{N-1}}\int d^{(N-1)}k \, e^{\mathrm{i} {\bf k}{\bf {\cdot }} \Delta {\bf x} } \sum_{j=0}^{n} (\Omega^{-1}_{{\bf k}})^{(j)}. \label{green-h1}
                \eea
For future convenience, we define the following (adiabatic) momentum integral 
\be
I^{(n)}_{\boldsymbol{ N}}=\frac{1}{2 (2\pi)^{N-1} a^{N-1}}\int d^{(N-1)}k \, e^{\mathrm{i} {\bf k} {\bf {\cdot}}\Delta {\bf x} }  (\Omega^{-1}_{{\bf k}})^{(n)}\, .
\ee
These integrals will be required in the following sections, where we explicitly compute the adiabatic expansion of the Feynman Green's function (\ref{green-h1}) in two, three and four space-time dimensions. We will then compare the resulting expressions with the DeWitt-Schwinger representation of the Feynman Green's function for spatial point-splitting. 
In $N$ space-time dimensions, we will work to adiabatic order $N$, sufficient for the renormalization of the SET.

\section{$N=2$}
\label{sec:two}
\subsection{Adiabatic expansion}
 For $N=2$ the scalar curvature is $R=\frac{2\ddot a}{a}$ and therefore the quantity $\chi $ (\ref{eq:chi}) takes the form
 \be
 \chi(t)=\left(2\xi -\frac{1}{2}\right)\frac{\ddot a}{a}+\frac{\dot a ^2}{4 a^2}.
 \ee 
 Since we only have one space dimension, for this section only we use the notation $k_1=k \in \mathbb R$. In this case we need to study the adiabatic expansion up to and including the second adiabatic order, namely
  \bea
               -\mathrm{i} G^{(2)}_{\textrm{Ad}}(t,x\, ; t, x^{\prime })=\frac{1}{4\pi a}\int_{-\infty}^{\infty} dk \, e^{\mathrm{i} k \Delta x} \Big((\Omega^{-1}_{{\bf k}})^{(0)}+(\Omega^{-1}_{{\bf k}})^{(1)}+(\Omega^{-1}_{{\bf k}})^{(2)}\Big).  \label{eq:greenAd-2D}
                \eea
                The momentum integrals can be easily performed with the aid of {\tt {MATHEMATICA}} software. For the leading and first orders we find
                \label{eq:In2}
                \begin{align} 
                \label{eq:I0n2}
              I^{(0)}_{{\bf 2}} & =  \frac{1}{4\pi a }\int_{-\infty}^{\infty} dk \, \frac{e^{\mathrm{i} k \Delta x}}{\omega}=\frac{1}{4\pi a }\Big(2 a K_0(m a \epsilon)\Big),
                \\ \label{eq:I1n2}
                 I^{(1)}_{{\bf2}} & =  \frac{1}{4\pi a }\int_{-\infty}^{\infty} dk \, \frac{e^{\mathrm{i} k \Delta x}(k q A)}{a^2\omega^3}=\frac{1}{4\pi a }\Big(2 \mathrm{i} q  A a \,\Delta x \, K_0(m a \epsilon)\Big),
                \end{align}
                where $K_j(z)$ are modified Bessel functions of  the second kind and $\epsilon=|\Delta x|$. For the second order integral we find  
                \begin{align} 
                I^{(2)}_{{\bf2}} &  =\frac{1}{4\pi a }\int_{-\infty}^{\infty} dk \, e^{\mathrm{i} k \Delta x}(\Omega_{{\bf k}}^{-1})^{(2)} 
                \nonumber \\ &  =\frac{1}{4\pi a }\Bigg(\Big[-a q^2 A^2 \epsilon^2 +\frac{1}{6}\epsilon^2 a^2 \ddot a\Big]K_0(ma \epsilon)
+  \Big[-\frac{1}{12}m \epsilon^2a^2\dot a^2+\frac{2}{m}(\tfrac{1}{6}-\xi)a \ddot a\,\Big]\epsilon\, K_1(m a \epsilon)\Bigg).
               \label{eq:I2n2}
                \end{align}
 The adiabatic expansion of the Feynman Green's function $ -\mathrm{i} G^{(2)}_{\textrm{Ad}}(t,x;t,x^{\prime} )$ is given by the sum of these three contributions. To compare this expansion with the Hadamard representation of the Feynman Green's function, we require a short-distance expansion of the adiabatic expansion \eqref{eq:greenAd-2D}. We have to consider terms up to order $O(\epsilon^2)$. Recall the expansions for small $z$ of the modified Bessel functions \cite{Abramowitz01}
                \begin{subequations}
                \label{Bessels}
                \bea \label{eq:K0exp}
                K_0(z)&\sim& -\gamma - \ln\left(\frac{z}{2}\right)+\frac{z^2}{4}\left[1-\gamma - \ln\left(\frac{z}{2}\right)\right] +...\\
                \label{eq:K1exp} K_1(z)&\sim& \frac{1}{z}+\frac{z}{2}\left[-\frac{1}{2}+\gamma + \ln\left(\frac{z}{2}\right)\right]+...
                \eea
                \end{subequations}
Introducing the expansions above in \eqref{eq:I2n2}, we obtain the short-distance expansion of the adiabatic expansion of the Feynman Green's function, which is
  \bea  
  -\mathrm{i} G^{(2)}_{\textrm{Ad}}(t,x\, ; t, x^{\prime })&=& \frac{1}{4\pi}\left[-1 -\mathrm{i}q A \Delta x + \frac{\epsilon^2}{2}\left(q^2 A^2-\frac{a^2m^2}{2}-\xi a \ddot a\right) \right]\left[\ln \left(\frac{\epsilon^2m^2 a^2}{4}\right)+2\gamma \right]
  \nonumber\\
  &&\quad + \frac{1}{4\pi}\Big(-\frac{2(\xi - \tfrac{1}{6})\ddot a}{m^2 a} + \frac{\epsilon^2}{2}\Big[a^2 m^2 + (\xi-\tfrac{1}{6}) a \ddot a \Big] - \frac{\dot a^2}{12}\epsilon ^2 \Big) + {\mathcal {O}}(\epsilon^3). 
  \label{eq:D2-final-adiabatic}
  \eea
We note that this expression is singular in the massless limit $m^{2}\to 0$, and contains, as well as terms depending logarithmically on $m^{2}$, terms of order $m^{-2}$ but no smaller powers of $m$.
Thus the powers of $m$ in (\ref{eq:D2-final-adiabatic}) match those in the DeWitt-Schwinger representation of the Green's function.  We now investigate whether the expansion (\ref{eq:D2-final-adiabatic}) is identical to the corresponding DeWitt-Schwinger (and thereby Hadamard) form of the Green's function.
  
  \subsection{Hadamard/DeWitt-Schwinger expansion}
  
  Recall that in $N=2$ space-time dimensions, the Hadamard representation of the Feynman Green's function reads
  \be \label{eq:Hadamard-d20}
  -\mathrm{i} G_{\mathrm{F}}(x, x^{\prime} )=\frac{1}{4 \pi}\left\{V(x, x^{\prime}) \ln \left[\frac{\sigma(x, x^{\prime} )}{\ell^{2}}+\mathrm{i} \varepsilon\right]+W(x, x^{\prime})\right\} \, .
  \ee
As discussed in Section \ref{sec:Hadamard},  $V(x,x^{\prime })$ and $W(x,x^{\prime })$ admit power series expansions in the geodesic interval \eqref{eq:powersH}, and the coefficients of the expansion further admit asymptotic  series expansions \eqref{eq:taylorUVW}. 
For $N=2$ the relevant terms (for renormalization) of the expansion are \cite{Balakumar:2019djw}
\begin{subequations}
\bea
V_0(x,x^{\prime })&=&V_{00}(x)+V_{01\mu}(x)\sigma^{;\mu}+V_{02\mu\nu}(x)\sigma^{;\mu}\sigma^{;\nu}+\mathcal{O}(\sigma^{3/2})\, ,\\
V_1(x,x^{\prime })&=&V_{10}(x)+\mathcal{O}(\sigma^{1/2})\, ,
\eea
where
\begin{align}
&V_{00}=-1\, ; \\
&V_{01 \mu}=-\mathrm{i} q A_{\mu}\, ; \\
&V_{02 \mu \nu}=-\frac{1}{24} R g_{\mu \nu}+\frac{\mathrm{i} q}{2} D_{(\mu} A_{\nu)}\, ,
\end{align}
and 
\be 
V_{10}=-\frac{1}{2}\left[m^{2}+\left(\xi-\frac{1}{6}\right) R\right]\, .
\ee
\end{subequations}
For our particular configuration, namely a flat FLRW space-time \eqref{FLRWmetric} and an electric field background described by the vector potential $A_\mu=(0,A(t))$, the short-distance expansion of the geometric term in \eqref{eq:Hadamard-d20} for spatially split points ($\Delta x\neq 0$ but $\Delta t=0$) reads 
\be \label{eq:D2-final-hadamard}
V(x,x^{\prime })\ln\left(\frac{\sigma}{\ell^2}\right)
=\frac{1}{4\pi}\left[-1-\mathrm{i} q A \Delta x+\frac{1}{2}\left(q^2 A^2-\frac{1}{2}a^2m^2-\xi a \ddot a \right)\epsilon ^2+ \mathcal O(\epsilon^3)\right]\ln\left(\frac{a^2\epsilon ^2}{2\ell^2}\right)  -\frac{\dot a^2}{48 \pi}\epsilon^2+ \mathcal O(\epsilon^3)\, ,
\ee
where we have used the short-distance expansion of the geodesic distance $\sigma$ derived in Appendix \ref{ap:geodesic-dist} [see Eq.~\eqref{eq:ansatz-sigma}] with $\Delta t=0$,
\be
\sigma =  \frac{a^2}{2}\epsilon^2 + \frac{a^2 \dot a ^2}{24}\epsilon^4  + \mathcal O(\epsilon^6)\, .
\ee
Comparing Eqs.~\eqref{eq:D2-final-adiabatic} and \eqref{eq:D2-final-hadamard} we see that the adiabatic and Hadamard short-distance expansions of the Green's function coincide except for terms which are either finite or vanish in the limit $\epsilon \to 0$.

We now compute the finite contribution of the DeWitt-Schwinger representation of the Feynman Green's function. Using the results \eqref{eq:VnAn-N2} we easily find
\begin{subequations}
\bea
\widetilde{\mathcal{A}}_0(m^2;x,x^{\prime })&=&-V_0(x,x^{\prime })\, ,\\
\widetilde{\mathcal{A}}_1(m^2;x,x^{\prime })&=&2V_1(x,x^{\prime })\, .
\eea
\end{subequations}
We use these results to find the mass-dependent DeWitt-Schwinger coefficient ${\widetilde {\mathcal {A}}}_{1}(m^{2};x,x^{\prime })$ and hence ${\mathcal {A}}_{1}(x,x^{\prime})$ using \eqref{eq:AnhatAn},
where the DeWitt-Schwinger coefficient has a covariant asymptotic  series expansion of the form
\be
\mathcal A_1(x,x^{\prime })=\mathcal A_{10}(x)+\mathcal A_{11\mu}(x)\sigma^{;\mu}+\mathcal A_{12\mu \nu}(x)\sigma^{;\mu} \sigma^{;\nu}+\mathcal{O}(\sigma^{3/2})\, .
\ee
Since evaluating the SET involves taking two derivatives of the Feynman Green's function, to renormalize this quantity we require the coefficients $W_0(x,x^{\prime })$ and $W_1(x,x^{\prime })$, given by \eqref{eq:Wn-N2}
\begin{subequations}
\bea
W_0(x,x^{\prime })&=&\left[2 \gamma+\ln\left( \frac{m^2\ell ^{2}}{2}\right) \right] V_{0}(x,x^{\prime }) + \frac{\mathcal{A}_1(x,x^{\prime })}{m^2}+\mathcal{O}(m^{-4})\, ,\\
W_1(x,x^{\prime })&=&\left[-2+2 \gamma+\ln\left( \frac{m^2\ell ^{2}}{2}\right) \right] V_1(x,x^{\prime }) + \frac{\mathcal{A}_1(x,x^{\prime })}{2}
+\mathcal{O}(m^{-2})\, .
\eea
\end{subequations}
Therefore, the short-distance expansion of the relevant terms of $W(x,x^{\prime })$ in our particular configuration reads 
\bea 
W(x,x^{\prime })&=&\frac{1}{4\pi}\left[ -1 -\mathrm{i}q A \Delta x + \frac{\epsilon^2}{2}\left(q^2 A^2-\frac{a^2m^2}{2}-\xi a \ddot a\right) \right]\left[\ln\left(\frac{m^2\ell ^{2}}{2}\right)+2\gamma\right] 
  \nonumber\\
  &&+\frac{1}{4\pi}\Big(-\frac{2(\xi - \tfrac{1}{6})\ddot a}{m^2 a} + \frac{\epsilon^2}{2}\Big[a^2 m^2 + (\xi-\tfrac{1}{6}) a \ddot a \Big] \Big) + {\mathcal {O}}(\epsilon^3). \label{eq:D2-finalW}
\eea
Combining \eqref{eq:D2-final-hadamard} and \eqref{eq:D2-finalW}, we find the short-distance expansion of the DeWitt-Schwinger representation of the Green's function, which does not depend on the renormalization length scale $\ell $. Furthermore, the resulting expression matches the adiabatic result \eqref{eq:D2-final-adiabatic}.  

\section{$\mathbf{N=3}$}
\label{sec:three}
\subsection{Adiabatic expansion}

For $N=3$ we have, from \eqref{eq:chi}, since $R=\tfrac{2{\dot {a}}^2}{a^{2}}+\frac{4{\ddot {a}}}{a}$, 
\be
\chi(t)=\frac{2\xi \dot a^2}{a^2}+(4\xi-1)\frac{\ddot a}{a}.
\ee
In this case, we want to compute
                \be
               -\mathrm{i} G^{(3)}_{\textrm{Ad}}(t,{\bf x}\, ; t, {\bf x^{\prime }})=\frac{1}{2(2\pi)^2 a^2}\int d^2k \, e^{\mathrm{i} {\bf k} {\bf {\cdot}}\Delta{\bf  x}} \Big((\Omega^{-1}_{{\bf k}})^{(0)}+(\Omega^{-1}_{{\bf k}})^{(1)}+(\Omega^{-1}_{{\bf k}})^{(2)}+(\Omega^{-1}_{{\bf k}})^{(3)}\Big)\, .  \label{eq:greenAd-3D}
               \ee
     As in the two-dimensional case, the momentum integrals can be performed using {\tt {MATHEMATICA}}. For the leading adiabatic order we find
     \be \label{eq:I0D3}
     I_{{\bf3}}^{(0)}=\frac{1}{2(2\pi)^2 a^2}\int d^2k \, \frac{e^{\mathrm{i} {\bf k}{\bf {\cdot}} \Delta{\bf  x}}}{\omega}=\frac{1}{4 \pi a^2}\int_{0}^{\infty} dk\, k \frac{J_0(k \epsilon)}{\omega}=\frac{e^{-m  a \epsilon}}{4\pi } \frac{  1}{a \epsilon},
     \ee
     where $J_{0}(k\epsilon )$ is a Bessel function of the first kind and we have used the result
     \be
     \int_{0}^{2\pi}e^{i k \epsilon \cos\theta}d\theta=2\pi J_0(k\epsilon) .
     \ee
     For the first adiabatic order, the integral is more subtle because it explicitly depends on $k_1$. However, we can use the following identity
     \be \label{eq:integral-trick-3D}
      \int d^2k  \, \frac{k_j \,e^{\mathrm{i} {\bf k}{\bf {\cdot }}\Delta {\bf x}}}{\omega^n}=-\mathrm{i}\partial_{x_j}\int d^2k \, \frac{e^{\mathrm{i}{\bf k}{\bf {\cdot }}\Delta {\bf x}}}{\omega^n}\, . 
     \ee
The left-hand side of the equation above is the integral in which we are interested, and the integral of the right-hand side can be easily computed using polar coordinates and then differentiated with respect to $x_j$. With this result, we can systematically obtain the nonhomogeneous integrals appearing in the adiabatic expansion from homogeneous integrals. Using this method, the adiabatic order one contribution reads
     \be
     I^{(1)}_{{\bf3}}=\frac{\mathrm{i}e^{-m a \epsilon}}{4\pi} \frac{ \Delta x}{a \epsilon}q A.
     \ee
     The second adiabatic order can be obtained in the same way. For the homogeneous integrals our computations are based on \eqref{eq:I0D3}, while, for the nonhomogeneous integrals where $k_1$ explicitly appears we use \eqref{eq:integral-trick-3D} and similar expressions. After some algebra we find
     \be
     I^{(2)}_{{\bf3}}=\frac{e^{-ma\epsilon}}{4 \pi}\left[\frac{1}{m}\left(\frac{1}{6}-\xi + \frac{m a \epsilon}{24}\right)\left(\frac{\dot a^2}{a^2}+ 2\frac{\ddot a}{a}\right)-\frac{m \epsilon^2  \dot a^2}{24}-\frac{ \Delta x^2 }{2a \epsilon }q^2 A^2\right].
     \ee
Similarly, the third adiabatic order gives 
     \be
     I^{(3)}_{{\bf 3}}=\frac{\mathrm{i} e^{-m a \epsilon}}{4 \pi}\left[\frac{1}{m}\left(\frac{1}{6}-\xi +\frac{m a \epsilon}{24}\right)\left(\frac{\dot a^2}{a^2}+2 \frac{\ddot a}{a}\right)\Delta x q A-\frac{m\epsilon^2\Delta x}{24}\dot a^2 q A+\frac{\epsilon \Delta x}{12}  \dot a \dot A -\frac{\Delta x^3}{6 a \epsilon}q^3 A^3-\frac{\Delta x}{12 m}
     q \ddot A\right].
     \ee
Using the short-distance expansion of the exponential, we easily arrive at the short-distance expansion of the (third order) adiabatic expansion of the Feynman Green's function
                \bea \label{eq:D3-final-adiabatic}
-\mathrm{i} G^{(3)}_{\textrm{Ad}}(t,{\bf x}\, ; t, {\bf x^{\prime }}) & = & \frac{1}{4 \pi \epsilon a}  +\frac{\mathrm{i} q A \Delta x}{4 a
   \pi  \epsilon}-\frac{m}{4 \pi }+\frac{\dot{a}^2}{24 a^2 m
   \pi }-\frac{\xi  \dot{a}^2}{4 a^2 m
   \pi }+\frac{\ddot{a}}{12 a m \pi
   }-\frac{\xi  \ddot{a}}{2 a m \pi }  -\frac{\mathrm{i} q A m \Delta x}{4 \pi
   }-\frac{q^2 A^2  \Delta x^2}{8 a \pi 
   \epsilon}
   \nonumber \\ & & 
   +\frac{\mathrm{i} q A  \dot{a}^2 \Delta x }{24 a^2
   m \pi }-\frac{\mathrm{i} q A   \xi 
   \dot{a}^2\Delta x}{4 a^2 m \pi }+\frac{{\mathrm {i}}q A
   q  \ddot{a}\Delta x}{12 a m \pi }
   -\frac{\mathrm{i} q A   \xi  \ddot{a} \Delta x}{2 a m
   \pi }  -\frac{\mathrm{i}   q \ddot{A}\Delta x}{48 m
   \pi }
   +\frac{a
   m^2 \epsilon}{8 \pi }-\frac{
   \dot{a}^2\epsilon}{32 a \pi }+\frac{ \xi 
   \dot{a}^2\epsilon}{4 a \pi }
   \nonumber \\ & &
   -\frac{
   \ddot{a}\epsilon}{16 \pi }+\frac{ \xi 
   \ddot{a}\epsilon}{2 \pi } +\frac{\mathrm{i} a q A
   m^2  \epsilon\Delta x}{8 \pi } -\frac{\mathrm{i} q A \dot{a}^2 \epsilon \Delta x
   }{32 a \pi } +\frac{\mathrm{i} q A  \xi 
   \dot{a}^2 \epsilon\Delta x}{4 a \pi }-\frac{\mathrm{i} q A  \ddot{a}\epsilon \Delta x}{16
   \pi }  +\frac{\mathrm{i} q A  \xi  \ddot{a}\epsilon \Delta x}{2 \pi
   } 
   \nonumber \\ & &
   +\frac{\mathrm{i}  \dot{a} q \dot{A} \epsilon \Delta x}{48
   \pi }+\frac{\mathrm{i} a q  \ddot{A}\epsilon \Delta x}{48
   \pi }-\frac{\mathrm{i} q^3 A^3
   \Delta x^3}{24 a \pi  \epsilon} +\frac{q^2 A^2 m 
   \Delta x^2}{8 \pi }-\frac{a^2 m^3
   \epsilon^2}{24 \pi }-\frac{m \epsilon^2
   \xi  \dot{a}^2}{8 \pi }+\frac{a m
   \epsilon^2 \ddot{a}}{48 \pi }
   \nonumber \\ & &
   -\frac{a m
   \epsilon^2 \xi  \ddot{a}}{4 \pi } +\mathcal O(\epsilon^3)\, .\label{adiabaticO3}
 \eea
As in two dimensions, we see that the powers of $m$ in the adiabatic expansion (\ref{adiabaticO3}) are those found in the DeWitt-Schwinger representation of the Green's function, truncated to the order necessary for the renormalization of the SET.

   \subsection{Hadamard/De Witt-Schwinger expansion}
   For $N=3$, the Hadamard representation of the Feynman Green's function reads
   \be
-\mathrm{i} G_{\mathrm{F}}(x, x^{\prime} )=\frac{1}{4 \sqrt{2}\,  \pi}\left\{\frac{U(x, x^{\prime} )}{\left[\sigma(x, x^{\prime} )+\mathrm{i} \varepsilon\right]^{\frac{1}{2}}}+W(x, x^{\prime} )\right\}\, .
\ee
   As in two space-time dimensions, the biscalars $U(x,x^{\prime })$ and $W(x,x^{\prime })$ can be expanded as power series in the geodesic interval $\sigma$ \eqref{eq:powersH}. Furthermore, for short distances, the coefficients of these expansions admit covariant asymptotic series expansions \eqref{eq:taylorUVW}. 
 For the renormalization of the SET, the relevant coefficients of these expansions are \cite{Balakumar:2019djw}
 \begin{subequations}
\bea
U_0(x,x^{\prime })&=&U_{00}(x)+U_{01\mu}(x)\sigma^{;\mu}+U_{02\mu \nu}(x)\sigma^{;\mu}\sigma^{;\nu}+U_{03\mu \nu \lambda}(x)\sigma^{;\mu}\sigma^{;\nu}\sigma^{;\lambda}+\mathcal{O}(\sigma^2)\, ,\\
U_1(x,x^{\prime })&=&U_{10}(x)+U_{11\mu}(x)\sigma^{;\mu}+\mathcal{O}(\sigma)\, ,
\eea
 where 
  \begin{align}
U_{00}&=1 ; \\
U_{01 \mu}&=\mathrm{i} q A_{\mu} ; \\
U_{02 \mu \nu}&=\frac{1}{12} R_{\mu \nu}-\frac{\mathrm{i} q}{2} D_{(\mu} A_{\nu)} ; \\
U_{03 \mu \nu \lambda}&=-\frac{1}{24} R_{(\mu \nu ; \lambda)}+\frac{\mathrm{i} q}{6} D_{(\mu} D_{\nu} A_{\lambda)}+\frac{\mathrm{i} q}{12} A_{(\mu} R_{\nu \lambda)} \, ,
\end{align}
and
\begin{align}
U_{10}&=m^{2}+\left(\xi-\frac{1}{6}\right) R\,; \\
U_{11 \mu}&=-\frac{1}{2}\left(\xi-\frac{1}{6}\right) R_{; \mu}+\mathrm{i} q\left[m^{2}+\left(\xi-\frac{1}{6}\right) R\right] A_{\mu}-\frac{\mathrm{i} q}{6} \nabla^{\alpha} F_{\alpha \mu}\, .
\end{align}
\end{subequations}
For our particular configuration of flat FLRW space-time \eqref{FLRWmetric} and electromagnetic potential $A_\mu=(0,A(t),0)$ the short-distance expansion of the geometric part of the Hadamard parametrix reads
\bea 
\frac{1}{4 \sqrt{2}\,  \pi} \, \frac{U(x,x^{\prime })}{\sigma(x,x^{\prime })^{\frac{1}{2}}}&=&\frac{1}{4 \pi a \epsilon}+\frac{\mathrm{i} q A \Delta x}{4 \pi a  \epsilon } +  \frac{a m^2 \epsilon}{8 \pi }-\frac{q^2 A^2 \Delta x^2}{8 a \pi 
   \epsilon^2}-\frac{  \dot a^2\epsilon}{32 a \pi }+\frac{
   \xi  \dot a^2 \epsilon}{4 a \pi }-\frac{
   \ddot a \epsilon}{16 \pi }+\frac{ \xi 
   \ddot a \epsilon }{2 \pi }+ \frac{\mathrm{i} a A m^2 q \epsilon\Delta x}{8 \pi }-\frac{\mathrm{i} q^3 A^3 
   \Delta x^3}{24 a \pi  \epsilon}
   \nonumber \\ & &
   -\frac{ \mathrm{i} q A  \dot a ^2\epsilon \Delta x
   }{32 a \pi }+\frac{\mathrm{i} q A   \xi 
   \dot a^2 \epsilon\Delta x}{4 a \pi }+\frac{\mathrm{i} q 
   \dot a \dot A\epsilon \Delta x}{48 \pi }-\frac{\mathrm{i} q A
    \ddot a \epsilon \Delta x}{16 \pi }+\frac{\mathrm{i} q A  
   \xi  \ddot a \epsilon\Delta x}{2 \pi }+\frac{\mathrm{i} a q 
   \ddot A \epsilon \Delta x}{48 \pi } + \mathcal O(\epsilon^3)\, ,
   \label{eq:D3-final-hadamard}
\eea
where we have used the result 
\be
\frac{1}{(2\sigma)^{\frac{1}{2}}}=\frac{1}{a\epsilon}-\frac{\epsilon^2\dot a^2}{24 a}+\frac{7(\dot a ^4- 24a \dot a^2 \ddot a)\epsilon^3}{5760  a}+\mathcal O(\epsilon^4)\, .
\ee
Comparing Eqs.~\eqref{eq:D3-final-adiabatic} and \eqref{eq:D3-final-hadamard} we see that they differ only by terms which are either finite or vanishing in the limit $\epsilon \to 0$, so that the adiabatic and Hadamard expansions of the Green's function have the same short-distance singularity structure.

As in two dimensions, these finite or vanishing terms can be directly obtained from the DeWitt-Schwinger representation of the Feynman Green's function, which 
is a particular choice of the Hadamard representation, with the coefficients $W_n(x,x^{\prime })$ given by \eqref{eq:Wn-DWS-odd}. For $N=3$  the mass-dependent DeWitt-Schwinger coefficients ${\widetilde {\mathcal {A}}}_{n}(m^{2};x,x^{\prime})$ that we require are related to the corresponding Hadamard coefficients by \eqref{eq:UAtildeodd}
\begin{subequations}
\bea
&&\widetilde{\mathcal A}_0(m^2;x,x^{\prime })=U_0(x,x^{\prime }),\\
&&\widetilde {\mathcal A}_1(m^2;x,x^{\prime })=-U_1(x,x^{\prime })\, ,
\eea
\end{subequations}
and using \eqref{eq:AnhatAn} we can find the DeWitt-Schwinger coefficients $\mathcal A_0(x,x^{\prime })$ and $\mathcal A_1(x,x^{\prime })$. Therefore, from \eqref{eq:Wn-DWS-odd}, the short-distance expansions of the biscalars $W_0(x,x^{\prime })$ and $W_1(x,x^{\prime })$ read
\begin{subequations}
\bea \label{eq:W0n3}
W_0(x,x^{\prime })&=&-\sqrt{2} m\mathcal{A}_0(x,x^{\prime })+\frac{\mathcal{A}_1(x,x^{\prime })}{\sqrt{2}m}+\mathcal{O}(m^{-3})\, ,\\
W_1(x,x^{\prime })&=&-\frac{\sqrt 2}{3}m^3 \mathcal{A}_0(x,x^{\prime })+\frac{1}{\sqrt{2}} m\mathcal{A}_1(x,x^{\prime })+\mathcal{O}(m^{-1})\, ,
\eea
\end{subequations}
where we employ  the following short-distance expansion of the DeWitt-Schwinger coefficients:
\begin{subequations}
\bea
\mathcal{A}_0(x,x^{\prime })&=&\mathcal{A}_{00}(x)+\mathcal{A}_{01\mu}(x)\sigma^{;\mu} +\mathcal{A}_{02\mu \nu}(x)\sigma^{;\mu}\sigma^{;\nu}+\mathcal{A}_{03\mu \nu \lambda}(x)\sigma^{;\mu}\sigma^{;\nu}\sigma^{;\lambda}+\mathcal{O}(\sigma^2)\, ,\\
\mathcal{A}_1(x,x^{\prime })&=&\mathcal{A}_{10}(x)+\mathcal{A}_{11\mu}(x)\sigma^{;\mu} +\mathcal{O}(\sigma)\, .
\eea
\end{subequations}
Using the above results, the finite DeWitt-Schwinger contribution to the Hadamard expansion reads as follows:
\bea
\frac{1}{4 \sqrt{2}\,  \pi}\,W(x,x^{\prime })&=&-\frac{m}{4 \pi }+\frac{\dot{a}^2}{24 a^2 m \pi }-\frac{\xi  \dot{a}^2}{4
   a^2 m \pi }+\frac{\ddot{a}}{12 a m \pi }-\frac{\xi  \ddot{a}}{2 a m \pi
   } -\frac{\mathrm{i} q A m  \Delta x}{4 \pi }+\frac{\mathrm{i} qA 
   \Delta x \dot{a}^2}{24 a^2 m \pi }-\frac{{\mathrm {i}}q A  \Delta x \xi  \dot{a}^2}{4 a^2 m \pi }+\frac{{\mathrm {i}}q A  \Delta x
   \ddot{a}}{12 a m \pi }
   \nonumber \\ & &
   -\frac{\mathrm{i} qA  \Delta x \xi  \ddot{a}}{2 a
   m \pi }  -\frac{\mathrm{i} q \Delta x \ddot{A}}{48 m \pi }-\frac{a^2 m^3 \epsilon^2}{24 \pi }+\frac{q^2A^2 m \Delta x^2}{8 \pi }-\frac{m \epsilon^2 \xi  \dot{a}^2}{8 \pi }+\frac{a m \epsilon^2 \ddot{a}}{48 \pi }-\frac{a m \epsilon^2 \xi  \ddot{a}}{4 \pi
   }
   \nonumber \\ & &
   -\frac{\epsilon^2
   \dot{a}^4}{24 a^2 m \pi }+\frac{\epsilon^2 \xi  \dot{a}^4}{4 a^2 m \pi
   }
   +\frac{\epsilon^2 \dot{a} a^{(3)}}{24 m \pi }-\frac{\epsilon^2 \xi  \dot{a}
   a^{(3)}}{4 m \pi }+\mathcal O(\epsilon^3)\, .
   \label{eq:WN3}
\eea
All the terms in the first three lines in (\ref{eq:WN3}) match the remaining finite or vanishing terms in the adiabatic expansion \eqref{adiabaticO3}.
However, the four terms in the last line of \eqref{eq:WN3} do not appear in the adiabatic expansion \eqref{adiabaticO3}.
These terms are of adiabatic order four and therefore do not appear in the adiabatic expansion \eqref{adiabaticO3}, 
where we have only considered terms up and including the third adiabatic order. 
Furthermore, these terms do not depend on the electromagnetic potential, are present for a neutral scalar field, and do not appear to have been considered previously in the literature.

Therefore, in three space-time dimensions, we find that the DeWitt-Schwinger and adiabatic expansions of the Green's functions are not identical, although they have the same short-distance singularity structure.
They differ by terms of order $m^{-1}$, order $\epsilon ^{2}$ and adiabatic order four. 
These will make a finite contribution to the renormalized SET, and arise from terms in  $W(x,x^{\prime})$ proportional to $(\xi-\tfrac{1}{6})R_{;\mu}\sigma^{;\mu}$. In a locally covariant renormalization scheme, such finite terms will make a local, purely geometric contribution to the renormalized SET \cite{Hollands:2004yh}. Here the presence of these additional terms in the DeWitt-Schwinger expansion means that it is not clear whether adiabatic renormalization is a locally covariant renormalization scheme in three dimensions. One would need to examine all the adiabatic order four terms in the adiabatic expansion, and compare with the DeWitt-Schwinger expansion in order to address this question.\footnote{We thank the anonymous referee for their invaluable insight on this point.}

\section{$\mathbf{N=4}$}
\label{sec:four}
\subsection{Adiabatic expansion}
In $N=4$ dimensions, the scalar curvature in terms of the scale factor is $R=6\big(\frac{\dot a^2}{a^2}+\frac{\ddot a}{a}\big)$, and therefore, from \eqref{eq:chi} we have
\be 
\chi(t)=\left(6\xi-\frac{3}{4}\right)\frac{\dot a^2}{a^2}+\left(6 \xi-\frac{3}{2}\right)\frac{\ddot a}{a} .
\label{eq:chi4}
\ee
The adiabatic expansion of the Feynman Green's function up to and including the fourth adiabatic order is 
\be
-\mathrm{i} G^{(4)}_{\textrm{Ad}}(t,{\bf x}\, ; t, {\bf x^{\prime }})=\frac{1}{2(2\pi)^3 a^3}\int d^3k \, e^{\mathrm{i} {\bf k}{\bf {\cdot }} \Delta{\bf  x}} \Big((\Omega^{-1}_{{\bf k}})^{(0)}+(\Omega^{-1}_{{\bf k}})^{(1)}+(\Omega^{-1}_{{\bf k}})^{(2)}+(\Omega^{-1}_{{\bf k}})^{(3)}+(\Omega^{-1}_{{\bf k}})^{(4)}\Big)  .
\label{eq:greenAd-4D}
               \ee
We can systematically compute the momentum integrals using the same techniques as in two and three space-time dimensions. 
The terms that only depend on $|{\bf k}|=k$ can be easily computed in (standard) spherical coordinates. For example, for $n=0$ we use the result 
\be 
\int_0^{2 \pi}\int_0^\pi e^{i k \epsilon \cos(\theta) }\sin(\theta) \, d\theta \,  d\phi=4 \pi\frac{\sin(k\epsilon)}{k\epsilon}
\ee
to find
\be \label{eq:d4o0-adiabatic}
               I^{(0)}_{{\bf4}}=\frac{1}{2 (2 \pi)^3a^3}\int d^3 k \, \frac{e^{\mathrm{i} {\bf k}  {\bf {\cdot }}\Delta{\bf  x}}}{\omega}=\frac{1}{(2 \pi)^2a^3}\int dk\, k^2 \frac{\sin(k \epsilon)}{k \epsilon}\frac{1}{\omega}=\frac{1}{4 \pi^2}\frac{m K_1(m a\epsilon )}{a\epsilon }\, .
               \ee
Terms that have an explicit dependence on $k_1$ can be obtained from homogeneous integrals in the same way as in $N=3$, namely
                 \be \label{eq:integral-trick-4D}
      \int d^3k  \, \frac{k_j \,e^{\mathrm{i} {\bf k}{\bf {\cdot }}\Delta {\bf x}}}{\omega^n}=-\mathrm{i}\partial_{x_j}\int d^3k \, \frac{e^{\mathrm{i} {\bf k} {\bf {\cdot }}\Delta {\bf x}}}{\omega^n}\, ,
     \ee
    where the right-hand side of the equation is homogeneous in $k$ and can be computed in spherical coordinates. The adiabatic order one integral is then
    \be
I^{(1)}_{{\bf4}}=\frac{1}{ 4\pi^2}\Big(\mathrm{i} q A  \Delta x \frac{m\,K_1(ma\epsilon)}{a \epsilon}\Big)\, .
\ee
Similarly we can compute the subsequent adiabatic integrals. For $n=2$ we find
\begin{align}
I^{(2)}_{{\bf4}}=\frac{1}{4\pi^2 }\Big[ &\frac{1}{2}(1-6\xi)\left(\frac{\dot{a}^2}{a^2}+\frac{\ddot{a}}{a}\right)K_0(m a 
   \epsilon)-\frac{m^2 \epsilon^2 \dot a^2}{24}K_0(m a
   \epsilon) + \frac{m a\epsilon}{12} \left(\frac{\dot{a}^2}{a^2}+\frac{\ddot{a}}{a}\right) K_1(m a \epsilon)-\frac{\Delta x^2 q^2A^2}{2}\frac{m\, K_1(m a \epsilon)}{ a \epsilon}\Big] \, ,
   \end{align}
   and  for $n=3$ we have  
\bea
I^{(3)}_{{\bf4}}=&&\frac{\mathrm{i}  q A \Delta x }{4\pi^2 }\left[ \frac{1}{2}(1-6\xi)\left( \frac{\dot{a}^2}{a^2}+\frac{\ddot{a}}{a}\right)K_0( m a
   \epsilon)-\frac{m^2 \epsilon^2 \dot a^2}{24}K_0( m a
   \epsilon)+ \frac{m a\epsilon}{12} \left(\frac{\dot{a}^2}{a^2}+\frac{\ddot{a}}{a}\right) K_1(m a \epsilon)\right]  \nonumber\\
   &&+\frac{1}{4 \pi^2}\left[ \frac{\mathrm{i} \Delta x q \dot A}{12 }\frac{\dot a}{a}\left\{ m a \epsilon K_1(m a \epsilon)-K_0(m a \epsilon )\right\} -\frac{\mathrm{i} \Delta x q^3 A^3}{6}\frac{K_1(m a \epsilon)}{a \epsilon}-\frac{\mathrm{i}\Delta x q\ddot A}{12}K_{0}(m a\epsilon)\right] \, .
\eea
Using the same approach, we find the fourth adiabatic order integral $I^{(4)}_{{\bf 4}}$. The resulting expression is somewhat lengthy and is given in Appendix \ref{ap:detailsD4}. 

Using the short-distance expansion of the modified Bessel functions \eqref{Bessels}, we obtain the short-distance expansion of the adiabatic expansion of the four-dimensional Feynman Green's function as follows:
\bea
-  \mathrm{i} G^{(4)}_{\textrm{Ad}}(t,{\bf x}\, ; t, {\bf x^{\prime }}) &  =  & \frac{1}{2 (2 \pi)^2} \Bigg\{ \frac{2}{a^2 \epsilon^2 }+\frac{2 \textrm{i} q A \Delta x}{a^2 \epsilon^2 }-\frac{q^2 A^2  \Delta x^2}{a^2
   \epsilon^2}-\frac{\dot{a}^2}{3 a^2}+\frac{3 \xi 
   \dot{a}^2}{a^2}-\frac{\ddot{a}}{3 a}+\frac{3
   \xi  \ddot{a}}{a}+\frac{\dot{a}^4}{4 a^4 m^2}-\frac{3 \xi  \dot{a}^4}{a^4
   m^2}+\frac{9 \xi ^2 \dot{a}^4}{a^4 m^2}
   \nonumber \\ & & 
   +\frac{29 \dot{a}^2 \ddot{a}}{30 a^3 m^2}-\frac{17 \xi
    \dot{a}^2 \ddot{a}}{2 a^3 m^2}+\frac{18 \xi ^2 \dot{a}^2 \ddot{a}}{a^3
   m^2}+\frac{3 \ddot{a}^2}{20 a^2 m^2}-\frac{5 \xi  \ddot{a}^2}{2 a^2
   m^2}+\frac{9 \xi ^2 \ddot{a}^2}{a^2 m^2}+\frac{q^2 \dot{A}^2}{12 a^2
   m^2}-\frac{3 \dot{a} a^{(3)}}{10 a^2 m^2}+\frac{3 \xi  \dot{a}
   a^{(3)}}{2 a^2 m^2} 
   \nonumber \\ & & 
   -\frac{a^{(4)}}{10 a m^2}+\frac{\xi  a^{(4)}}{2 a
   m^2} +\frac{1}{12} \textrm{i} \Delta x q\ddot{A}-\frac{\textrm{i}q^3 A^3
   \Delta x^3}{3 a^2 \epsilon^2}+\frac{\textrm{i} \Delta x \dot{a} q \dot{A}}{4 a} -\frac{\textrm{i} q A \Delta x \dot{a}^2}{3 a^2} 
   +\frac{3 \textrm{i} q A \Delta  x \xi
    \dot{a}^2}{a^2}
    -\frac{\textrm{i} q A \Delta x \ddot{a}}{3 a}
    \nonumber \\ & &
    +\frac{3 \textrm{i} q A  \Delta x \xi 
   \ddot{a}}{a}  -\frac{3}{32} a^2 m^4 \epsilon^2+\frac{1}{6} m^2 \epsilon^2 \dot{a}^2+\frac{1}{8} a m^2 \epsilon^2 \ddot{a}-\frac{3}{4} m^2 \epsilon^2
   \xi  \dot{a}^2-\frac{3}{4} a m^2 \epsilon^2 \xi
    \ddot{a} +\frac{q^2 A^2 \Delta x^2 \dot{a}^2}{6
   a^2}
   \nonumber \\ & &
   -\frac{3 q^2 A^2 \Delta x^2 \xi  \dot{a}^2}{2 a^2} +\frac{q^2 A^2
   \Delta x^2 \ddot{a}}{6 a} -\frac{3 q^2 A^2 \Delta x^2 \xi  \ddot{a}}{2 a} -\frac{2 \epsilon^2
   \dot{a}^4}{45 a^2}+\frac{\epsilon^2 \xi  \dot{a}^4}{4 a^2}+\frac{5 \epsilon^2 \xi 
   \dot{a}^2 \ddot{a}}{4 a}-\frac{1}{40} \epsilon^2 \ddot{a}^2
   \nonumber \\ & &
   +\frac{1}{4} \epsilon^2
   \xi  \ddot{a}^2-\frac{127 \epsilon^2 \dot{a}^2 \ddot{a}}{720 a} -\frac{1}{80} \epsilon^2 \dot{a} a^{(3)}+\frac{1}{4} \epsilon^2 \xi 
   \dot{a} a^{(3)}+\frac{1}{240} a \epsilon^2 a^{(4)}+\frac{q^4 A^4  \Delta x^4}{12 a^2 \epsilon^2} -\frac{q^2 A  \dot{A} \dot{a} 
   \Delta x^2}{4 a}
   \nonumber \\ & &
   -\frac{1}{24} q^2 \Delta x^2 \dot{A}^2-\frac{1}{12} q^2 A\ddot{A}
    \Delta x^2   +\mathcal O(\epsilon^3) +  \Big[\ln \Big( \frac{\epsilon^2 m^2 a^2}{4}\Big)-1+2\gamma\Big] w_{\ln} \Bigg\} \, ,
    \label{eq:adiabatic4D}
\eea
where we have defined the quantity
\bea
w_{\ln} &  = &  \frac{m^2}{2}-\frac{\dot{a}^2}{2 a^2}+\frac{3 \xi 
   \dot{a}^2}{a^2}-\frac{\ddot{a}}{2 a}+\frac{3 \xi  \ddot{a}}{a} 
   +\frac{1}{2} \mathrm{i} q A m^2 \Delta x-\frac{\mathrm{i}q  A \dot{a}^2 \Delta x }{2 a^2}+\frac{3 \mathrm{i} q A \xi  \dot{a}^2 \Delta x }{a^2}-\frac{\mathrm{i} q A  \ddot{a}
   \Delta x }{2 a}
   +\frac{3 \mathrm{i} q A \xi  \ddot{a} \Delta x }{a}
   \nonumber \\ & & 
   +\frac{\mathrm{i} q 
   \dot{a} \dot{A}\Delta x}{12 a}+\frac{1}{12} \mathrm{i} q \ddot{A}\Delta x -\frac{1}{4} q^2A^2 m^2
   \Delta x^2
   +\frac{1}{16} a^2 m^4 \epsilon ^2+\frac{q^2 A^2 \dot{a}^2\Delta x^2 }{4 a^2}-\frac{1}{24}
   m^2 \epsilon ^2 \dot{a}^2-\frac{3q^2 A^2 \xi  \dot{a}^2 \Delta x^2 }{2 a^2}
   \nonumber \\ & & 
   +\frac{3}{4} m^2 \epsilon ^2 \xi 
   \dot{a}^2+\frac{\epsilon ^2 \dot{a}^4}{16 a^2}-\frac{3 \epsilon ^2 \xi  \dot{a}^4}{4 a^2}+\frac{9 \epsilon ^2 \xi ^2
   \dot{a}^4}{4 a^2}+\frac{A^2 q^2 \Delta x^2 \ddot{a}}{4 a}
   -\frac{1}{12} a m^2 \epsilon ^2 \ddot{a}-\frac{3
   q^2 A^2  \Delta x^2 \xi  \ddot{a}}{2 a}
   +\frac{3}{4} a m^2 \epsilon ^2 \xi  \ddot{a}
   \nonumber \\ & & 
   +\frac{\epsilon ^2
   \dot{a}^2 \ddot{a}}{16 a}
   -\frac{9 \epsilon ^2 \xi  \dot{a}^2 \ddot{a}}{8 a}+\frac{9 \epsilon ^2 \xi ^2 \dot{a}^2
   \ddot{a}}{2 a}-\frac{3}{8} \epsilon ^2 \xi  \ddot{a}^2+\frac{9}{4} \epsilon ^2 \xi ^2 \ddot{a}^2-\frac{q^2 A\dot{A} \dot{a}
   \Delta x^2  }{12 a}
   -\frac{1}{24} q^2 \dot{A}^2 \Delta x^2 +\frac{1}{48} q^2 \epsilon
   ^2 \dot{A}^2
   \nonumber \\ & & 
   -\frac{1}{12} q^2 A \ddot{A} \Delta x^2
   -\frac{5}{48} \epsilon ^2 \dot{a} a^{(3)}+\frac{5}{8}
   \epsilon ^2 \xi  \dot{a} a^{(3)}-\frac{1}{48} a \epsilon ^2 a^{(4)}+\frac{1}{8} a \epsilon ^2 \xi 
   a^{(4)}+\mathcal{O}(\epsilon^3) \, .
   \label{eq:VLOG-4D}
\eea
\subsection{Hadamard/DeWitt-Schwinger expansion}
The Hadamard expansion of the Feynman Green's function in $N=4$ space-time dimensions is given by 
\begin{equation}
\begin{aligned}
-\mathrm{i} G_{\mathrm{F}}(x, x^{\prime} )= \frac{1}{2 (2 \pi)^2} \left\{\frac{U(x, x^{\prime} )}{\sigma(x, x^{\prime} )+\mathrm{i} \varepsilon}
+V(x, x^{\prime} ) \ln \left[\frac{\sigma(x, x^{\prime} )}{\ell^{2}}+\mathrm{i} \varepsilon\right]+W(x, x^{\prime} )\right\} \, .
\end{aligned}
\end{equation}
The coefficients $V(x, x^{\prime} )$ and $W(x, x^{\prime} )$ can be expanded in powers of $\sigma(x, x^{\prime} )$ as detailed in \eqref{eq:powersH}.
In this case, the expansion of $U(x,x^{\prime })$ in terms of $\sigma$ only has one term, that is $U(x,x^{\prime })\equiv U_0(x,x^{\prime })$. The coefficients $V_n(x,x^{\prime })$, $U_0(x,x^{\prime })$ and $W_n(x,x^{\prime })$ admit covariant asymptotic  expansions \eqref{eq:taylorUVW}. The relevant terms for renormalization of the SET are
\begin{subequations}
\bea
U_0(x,x^{\prime })&=&U_{00}(x)+U_{01\mu}(x)\sigma^{;\mu}+U_{02\mu \nu}(x)\sigma^{;\mu}\sigma^{;\nu}+U_{03\mu \nu \lambda}(x)\sigma^{;\mu}\sigma^{;\nu}\sigma^{;\lambda}+U_{04\mu \nu \lambda \tau}(x)\sigma^{;\mu}\sigma^{;\nu}\sigma^{;\lambda}\sigma^{;\tau}+\mathcal{O}(\sigma^{5/2})\, , \qquad\\
V_0(x,x^{\prime })&=&V_{00}(x)+V_{01\mu}(x)\sigma^{;\mu}+V_{02\mu \nu}(x)\sigma^{;\mu}\sigma^{;\nu}+\mathcal{O}(\sigma^{3/2})\, ,\\
V_1(x,x^{\prime })&=&V_{10}(x)+\mathcal{O}(\sigma^{1/2})\, ,
\eea
where
\begin{align}
U_{00} &=1 ; \\
U_{01 \mu} &=\mathrm{i} q A_{\mu} ; \\
U_{02 \mu \nu} &=\frac{1}{12} R_{\mu \nu}-\frac{\mathrm{i} q}{2} D_{(\mu} A_{\nu)} ;\\
U_{03 \mu \nu \lambda} &=-\frac{1}{24} R_{(\mu \nu ; \lambda)}+\frac{\mathrm{i} q}{6} D_{(\mu} D_{\nu} A_{\lambda)}+\frac{\mathrm{i} q}{12} A_{(\mu} R_{\nu \lambda)} ; \\
U_{04 \mu \nu \lambda \tau} &=\frac{1}{80} R_{(\mu \nu ; \lambda \tau)}+\frac{1}{360} R^{\rho}{}_{(\mu|\psi| \nu} R^{\psi}{ }_{\lambda|\rho| \tau)}+\frac{1}{288} R_{(\mu \nu} R_{\lambda \tau)} 
-\frac{\mathrm{i} q}{24} D_{(\mu} D_{\nu} D_{\lambda} A_{\tau)}-\frac{\mathrm{i} q}{24} D_{(\mu}\left[A_{\nu} R_{\lambda \tau)}\right]\, ,
\end{align}
and in addition
\begin{align}
V_{00}=& \frac{1}{2}\left[m^{2}+\left(\xi-\frac{1}{6}\right) R\right] ; \\
V_{01 \mu}=&-\frac{1}{4}\left(\xi-\frac{1}{6}\right) R_{; \mu}+\frac{\mathrm{i} q}{2}\left[m^{2}+\left(\xi-\frac{1}{6}\right) R\right] A_{\mu} -\frac{\mathrm{i} q}{12} \nabla^{\alpha} F_{\alpha \mu} ; \\
V_{02 \mu \nu}=& \frac{1}{24}\left[m^{2}+\left(\xi-\frac{1}{6}\right) R\right] R_{\mu \nu}+\frac{1}{12}\left(\xi-\frac{3}{20}\right) R_{; \mu \nu}-\frac{1}{240} \square R_{\mu \nu} 
+\frac{1}{180} R^{\alpha}{}_{\mu} R_{\alpha \nu}-\frac{1}{360} R^{\alpha \beta} R_{\alpha \mu \beta \nu}
\nonumber \\ & 
-\frac{1}{360} R^{\alpha \beta \gamma}{ }_{\mu} R_{\alpha \beta \gamma \nu} 
-\frac{\mathrm{i} q}{4}\left[m^{2}+\left(\xi-\frac{1}{6}\right) R\right] D_{(\mu} A_{\nu)}
-\frac{\mathrm{i} q}{4}\left(\xi-\frac{1}{6}\right) A_{(\mu} R_{; \nu)} 
-\frac{q^{2}}{24} F^{\alpha}{ }_{\mu} F_{\nu \alpha}-\frac{q^{2}}{12} A_{(\mu} \nabla^{\alpha} F_{\nu) \alpha}
\nonumber \\ & 
-\frac{\mathrm{i} q}{24} \nabla_{(\mu} \nabla^{\alpha} F_{\nu) \alpha}\, ,
\end{align}
together with
\be
V_{10}=\frac{1}{8}\left[m^{2}+\left(\xi-\frac{1}{6}\right) R\right]^{2}-\frac{1}{24}\left(\xi-\frac{1}{5}\right) \square R-\frac{1}{720} R^{\alpha \beta} R_{\alpha \beta} 
+\frac{1}{720} R^{\alpha \beta \gamma \delta} R_{\alpha \beta \gamma \delta}-\frac{q^{2}}{48} F^{\alpha \beta} F_{\alpha \beta} \,.
\ee
\end{subequations}
The explicit computation of the short distance expansion using the metric \eqref{FLRWmetric} and $A_\mu=(0,A(t),0,0)$ results in
\bea 
\frac{U_0(x,x^{\prime })}{\sigma(x,x^{\prime })} & = &\frac{2}{a^2 \epsilon^2}+\frac{2 \mathrm{i} q A  \Delta x}{a^2 \epsilon^2 }-\frac{q^2 A^2  \Delta x^2}{a^2
   \epsilon^2}+\frac{\dot{a}^2}{6 a^2}+\frac{\ddot{a}}{6 a}-\frac{\mathrm{i} q^3 A^3 \Delta x^3}{3 a^2 \epsilon^2}+\frac{\mathrm{i} q A
   \Delta x \dot{a}^2}{6 a^2}+\frac{\mathrm{i} q A  \Delta x \ddot{a}}{6 a}+\frac{\mathrm{i} q 
   \dot{a} \dot{A}\Delta x}{6 a}+ \frac{q^4 A^4 \Delta x^4}{12 a^2 \epsilon^2}
   \nonumber \\ & & 
   -\frac{q^2A^2 \Delta x^2 \dot{a}^2}{12 a^2}-\frac{\epsilon^2 \dot{a}^4}{360 a^2}-\frac{q^2A^2 \Delta x^2 \ddot{a}}{12 a}+\frac{2 \epsilon^2
   \dot{a}^2 \ddot{a}}{45 a}+\frac{1}{80} \epsilon^2 \ddot{a}^2-\frac{q^2A \dot{A} \dot{a} \Delta x^2 }{6
   a}+\frac{1}{60} \epsilon^2 \dot{a} a^{(3)} + \mathcal O(\epsilon^3)\, ,
   \label{eq:U4D}
\eea
and
\be
V(x,x^{\prime })\ln\left(\frac{\sigma}{\ell^2}\right)=\frac{1}{24} m^2 \epsilon ^2 \dot{a}^2-\frac{\epsilon ^2 \dot{a}^4}{24 a^2}+\frac{\epsilon ^2 \xi 
   \dot{a}^4}{4 a^2}-\frac{\epsilon ^2 \dot{a}^2 \ddot{a}}{24 a}+\frac{\epsilon ^2 \xi  \dot{a}^2 \ddot{a}}{4
   a} +\mathcal O(\epsilon^3)+w_{\ln} \ln \left(\frac{a^2 \epsilon ^2}{2\ell^2}\right) \, ,
\label{eq:V4D}
\ee
where $w_{\ln }$ is given by \eqref{eq:VLOG-4D}.
We now combine (\ref{eq:U4D}, \ref{eq:V4D}) and compare the result with the adiabatic expansion \eqref{eq:adiabatic4D}.
As anticipated, all singular terms in (\ref{eq:U4D}, \ref{eq:V4D}) exactly match those in \eqref{eq:adiabatic4D}, but there are terms in \eqref{eq:adiabatic4D} which are either finite or vanishing as $\epsilon \to 0 $ and which do not appear in the Hadamard parametrix.  

To account for these additional terms, we turn to the DeWitt-Schwinger representation of the Feynman Green's function, obtained from Eq. \eqref{eq:DWSn-even}, giving: 
\begin{subequations}
\bea
\widetilde{\mathcal{A}}_0(m^2;x,x^{\prime })&=&U_0(x,x^{\prime }) \, ,\\
\widetilde{  \mathcal{A}}_1(m^2;x,x^{\prime })&=&-2 V_0(x,x^{\prime }) \,  ,\\
\widetilde{  \mathcal{A}}_2(m^2;x,x^{\prime })&=&4 V_1(x,x^{\prime }) \, ,
\eea
\end{subequations}
and
\begin{subequations}
\bea
W_0(x,x^{\prime })&=&\left[ -1+2\gamma+\ln\left( \frac{m^2\ell ^{2}}{2} \right) \right] V_0(x,x^{\prime })-\frac{\mathcal{A}_1(x,x^{\prime })}{2}+\frac{\mathcal{A}_2(x,x^{\prime })}{2m^2}+\mathcal{O}(m^{-4}) \, ,\\
W_1(x,x^{\prime })&=&\left[-\frac{5}{2}+2\gamma+\ln\left(  \frac{m^2\ell ^{2}}{2} \right)  \right]  V_1(x,x^{\prime })-\frac{m^2\mathcal{A}_1(x,x^{\prime })}{8}+\frac{3\mathcal{A}_2(x,x^{\prime })}{8}+\mathcal{O}(m^{-2})\, .
\eea
\end{subequations}
For the short-distance expansions of the DeWitt-Schwinger coefficients we use 
\begin{subequations}
\bea 
\mathcal A_1(x,x^{\prime })&=&\mathcal{A}_{10}(x)+\mathcal{A}_{11\mu}(x)\sigma^{; \mu} +\mathcal{A}_{12\mu \nu}(x)\sigma^{; \mu}\sigma^{; \nu} + \mathcal O(\sigma^{3/2})\, ,\\
{\mathcal {A}}_2(x,x^{\prime })&=&\mathcal{A}_{20}(x) + \mathcal O(\sigma^{1/2})\, ,
\eea 
\end{subequations}
as required for the renormalization of the SET. A lengthy but straightforward calculation then gives  the short-distance expansion of the coefficient $W(x,x^{\prime })$ to be 
\bea
W(x,x^{\prime }) &  = & -\frac{\dot{a}^2}{2 a^2}+\frac{3 \xi  \dot{a}^2}{a^2}-\frac{\ddot{a}}{2 a}+\frac{3 \xi  \ddot{a}}{a}+\frac{29 \dot{a}^2 \ddot{a}}{30
   a^3 m^2}+\frac{\dot{a}^4}{4 a^4
   m^2}-\frac{3 \xi  \dot{a}^4}{a^4 m^2}+\frac{9 \xi ^2 \dot{a}^4}{a^4
   m^2}-\frac{17 \xi  \dot{a}^2 \ddot{a}}{2 a^3 m^2}+\frac{18 \xi ^2 \dot{a}^2
   \ddot{a}}{a^3 m^2}+\frac{3 \ddot{a}^2}{20 a^2 m^2}
   \nonumber \\ & & 
   -\frac{5 \xi  \ddot{a}^2}{2 a^2
   m^2}+\frac{9 \xi ^2 \ddot{a}^2}{a^2 m^2}-\frac{3
   \dot{a} a^{(3)}}{10 a^2 m^2} +\frac{3 \xi  \dot{a} a^{(3)}}{2 a^2
   m^2}-\frac{a^{(4)}}{10 a m^2}+\frac{\xi  a^{(4)}}{2 a m^2}+\frac{q^2 \dot{A}^2}{12 a^2 m^2} +\frac{1}{12} \textrm{i}  \Delta x q \ddot{A}-\frac{\textrm{i} q A \Delta x \dot{a}^2}{2 a^2}
   \nonumber \\ & & 
   +\frac{3 \textrm{i} q A \Delta x  \xi 
   \dot{a}^2}{a^2}-\frac{\textrm{i} q A \Delta x \ddot{a}}{2 a}
   +\frac{3 \textrm{i} q A 
   \Delta x \xi  \ddot{a}}{a}+\frac{\textrm{i}  \Delta x  \dot{a} q\dot{A}}{12
   a}-\frac{3}{32}
   a^2 m^4 \epsilon^2+\frac{1}{8} m^2 \epsilon^2 \dot{a}^2+\frac{1}{8} a m^2 \epsilon^2 \ddot{a}-\frac{3}{4} m^2 \epsilon^2 \xi  \dot{a}^2
   \nonumber \\ & & 
   -\frac{3}{4} a m^2 \epsilon^2 \xi  \ddot{a}+\frac{q^2 A^2  \Delta x^2
   \dot{a}^2}{4 a^2}-\frac{3 q^2 A^2  \Delta x^2 \xi  \dot{a}^2}{2 a^2}+\frac{q^2 A^2 
   \Delta x^2 \ddot{a}}{4 a}-\frac{3 q^2 A^2
    \Delta x^2 \xi  \ddot{a}}{2 a}-\frac{43 \epsilon^2 \dot{a}^2 \ddot{a}}{240
   a}+\frac{\epsilon^2 \xi  \dot{a}^2 \ddot{a}}{a}-\frac{3}{80} \epsilon^2 \ddot{a}^2
   \nonumber \\ & & 
   +\frac{1}{4}
   \epsilon^2 \xi  \ddot{a}^2-\frac{7}{240} \epsilon^2 \dot{a} a^{(3)}+\frac{1}{4} \epsilon^2 \xi  \dot{a}
   a^{(3)} +\frac{1}{240} a \epsilon^2 a^{(4)}-\frac{1}{12}   \Delta x^2 q^2 A\ddot{A}-\frac{1}{24}  \Delta x^2 q^2 \dot{A}^2 -\frac{  \Delta x^2 \dot{a} q^2 A  \dot{A}}{12
   a}+\mathcal{O}(\epsilon^3) 
   \nonumber \\ & & 
   + \Big[\ln \Big(\frac{m^2\ell ^{2}}{2}\Big) -1+2 \gamma\Big]w_{\ln}\, ,
   \label{eq:W4D}
\eea
where $w_{\ln}$ is given in  \eqref{eq:VLOG-4D}.
Combining (\ref{eq:U4D},  \ref{eq:V4D}, \ref{eq:W4D}), the renormalization length scale $\ell $ drops out of the resulting expression, and the DeWitt-Schwinger Green's function reproduces precisely the adiabatic expansion of the Feynman Green's function (\ref{eq:adiabatic4D}).

\section{Conclusions}
\label{sec:conc}

We have studied three approaches to the renormalization of expectation values of operators acting on a charged quantum scalar field: Hadamard, DeWitt-Schwinger and adiabatic. 
For any number of space-time dimensions greater than or equal to two, we have explicitly demonstrated that the DeWitt-Schwinger representation of the Feynman Green's function has the Hadamard form, generalizing the result of Ref.~\cite{Decanini:2005gt} for a neutral scalar field. 
The DeWitt-Schwinger Green's function corresponds to a particular choice of the Hadamard coefficient $W(x,x^{\prime })$ depending on the background geometry, electromagnetic potential and scalar field parameters. 
This coefficient is  set equal to zero in Hadamard renormalization.
As a result, while both Hadamard and DeWitt-Schwinger renormalization yield finite expectation values of the scalar field condensate, charge current and SET, the resulting renormalized SET components will differ by a local, conserved, geometric tensor, in accordance with Wald's axioms \cite{Wald:1977up}.

Specializing to a background consisting of a spatially-flat FLRW universe in two, three and four dimensions, with a time-dependent electromagnetic potential, we calculated the adiabatic expansion of the Feynman Green's function with spatial point-splitting, working to the order required for the renormalization of the SET.
We also find expressions for the Hadamard and DeWitt-Schwinger representations in this case, performing an asymptotic series expansion in the spatial separation, again truncating the expansion once we have sufficient terms for a computation of the renormalized SET. 
The inclusion of a time-dependent background electric field makes our expressions considerably more complex than those for a neutral scalar. 

As anticipated, all three representations of the Feynman Green's function (Hadamard, DeWitt-Schwinger and adiabatic) have identical terms singular in the coincidence limit. However, the terms which are finite (or vanishing) as the spatial separation tends to zero do not always agree.
In two and four dimensions, the DeWitt-Schwinger and adiabatic Green's function match exactly, demonstrating that adiabatic renormalization is a locally covariant renormalization scheme for a charged scalar field in this case. 
In three dimensions, we find terms in the DeWitt-Schwinger Green's function which are not present in the adiabatic expansion. 
These terms are quadratic in the spatial separation and hence make a finite contribution to the renormalized SET. 
However, they are of higher adiabatic order than required for the renormalization of the SET, and hence have been ignored in the adiabatic expansion. 
Therefore, in three dimensions, it is not necessarily the case that adiabatic renormalization is a locally covariant renormalization scheme.

Our results provide strong evidence for the robustness of the adiabatic approach to renormalization for a charged scalar field in even dimensions. Assuming that the background electromagnetic field is of adiabatic order one, adiabatic renormalization gives the correct conformal anomaly \cite{FN:2018}. We have further demonstrated that the results for the renormalized scalar condensate, charged scalar current and SET obtained using the adiabatic approach are identical to those arising from DeWitt-Schwinger renormalization.
In odd dimensions, the situation is less clear. Further investigation of higher-order adiabatic terms is required in order to ascertain whether adiabatic renormalization is a locally covariant renormalization prescription.

\appendix

\section{$N=4$ details} \label{ap:detailsD4}

The adiabatic function $(\Omega_{{\bf k}}^{-1})^{(4)}$ is 
\bea
(\Omega_{{\bf k}}^{-1})^{(4)}& = &\frac{35k_1^4 q^4A^4  }{8 a^8 \omega ^9}-\frac{15k_1^2 q^4 A^4 
   }{4 a^6 \omega ^7}-\frac{15 k_1^2 q^2 A^2  \chi }{4 a^4
   \omega ^7}+\frac{3 q^4 A^4}{8 a^4 \omega ^5}+\frac{3 q^2 A^2 \chi }{4
   a^2 \omega ^5}+\frac{3 \chi^2}{8 \omega ^5} -\frac{315 m^4 q^2 k_1^2 A^2
      \dot{a}^2}{16 a^6 \omega ^{11}}+\frac{35 m^2
   k_1^2  q^2 A^2 \dot{a}^2}{4 a^6 \omega ^9}
   \nonumber \\ & & 
   +\frac{35 m^4 q^2 A^2 
   \dot{a}^2}{16 a^4 \omega ^9}+\frac{35 m^4 \chi  \dot{a}^2}{16 a^2
   \omega ^9} +\frac{15  k_1^2 q^2 A^2 \dot{a}^2}{16 a^6 \omega
   ^7}-\frac{5  m^2 q^2 A^2 \dot{a}^2}{4 a^4 \omega ^7}-\frac{5 m^2 \chi 
   \dot{a}^2}{2 a^2 \omega ^7}-\frac{3  q^2 A^2  \dot{a}^2}{16 a^4 \omega
   ^5}+\frac{5 \chi  \dot{a}^2}{16 a^2 \omega ^5} +\frac{1155 m^8
   \dot{a}^4}{128 a^4 \omega ^{13}}
   \nonumber \\ & & 
   -\frac{231 m^6 \dot{a}^4}{16 a^4 \omega
   ^{11}}+\frac{357 m^4 \dot{a}^4}{64 a^4 \omega ^9}-\frac{3 m^2
   \dot{a}^4}{16 a^4 \omega ^7}+\frac{3 \dot{a}^4}{128 a^4 \omega
   ^5} +\frac{35 m^2 k_1^2  q^2 A^2 \ddot{a}}{8 a^5 \omega
   ^9}-\frac{15  k_1^2 q^2 A^2 \ddot{a}}{8 a^5 \omega ^7}-\frac{5 m^2 q^2 A^2
    \ddot{a}}{8 a^3 \omega ^7}-\frac{5 m^2 \chi  \ddot{a}}{8 a
   \omega ^7}
   \nonumber \\ & & 
   +\frac{3 q^2 A^2  \ddot{a}}{8 a^3 \omega ^5} +\frac{5 \chi
   \ddot{a}}{8 a \omega ^5}-\frac{231 m^6 \dot{a}^2 \ddot{a}}{32 a^3
   \omega ^{11}}+\frac{315 m^4 \dot{a}^2 \ddot{a}}{32 a^3 \omega
   ^9}-\frac{81 m^2 \dot{a}^2 \ddot{a}}{32 a^3 \omega ^7}-\frac{3
   \dot{a}^2 \ddot{a}}{32 a^3 \omega ^5}+\frac{21 m^4 \ddot{a}^2}{32 a^2
   \omega ^9}-\frac{3 m^2 \ddot{a}^2}{4 a^2 \omega ^7}+\frac{3
   \ddot{a}^2}{32 a^2 \omega ^5}
   \nonumber \\ & & 
   +\frac{35  k_1^2 m^2  \dot{a}
   q^2 A \dot{A}}{4 a^5 \omega ^9}-\frac{5  k_1^2 q^2 A \dot{A}\dot{a} }{4
   a^5 \omega ^7}-\frac{5  m^2 \dot{a} q^2 A  \dot{A}}{4 a^3 \omega
   ^7} +\frac{ \dot{a} q^2 A \dot{A}}{4 a^3 \omega ^5}-\frac{5 k_1^2
   q^2 \dot{A}^2}{8 a^4 \omega ^7}+\frac{q^2 \dot{A}^2}{4 a^2 \omega
   ^5}-\frac{5  k_1^2 q^2 A \ddot{A}}{4 a^4 \omega ^7}+\frac{q^2 A
   \ddot{A}}{4 a^2 \omega ^5}
   \nonumber \\ & & 
   -\frac{5 m^2 \dot{a} \dot{\chi }}{8 a
   \omega ^7} +\frac{5 \dot{a} \dot{\chi }}{8 a \omega
   ^5}+\frac{\ddot{\chi }}{8 \omega ^5}+\frac{7 m^4 \dot{a} a^{(3)}}{8
   a^2 \omega ^9}-\frac{m^2 \dot{a} a^{(3)}}{a^2 \omega ^7}+\frac{\dot{a}
   a^{(3)}}{8 a^2 \omega ^5}-\frac{m^2 a^{(4)}}{16 a \omega
   ^7}+\frac{a^{(4)}}{16 a \omega ^5}\, ,
\eea
where $\chi $ is given by \eqref{eq:chi4}. 
The integral of fourth adiabatic order 
\be 
I^{(4)}_{{\bf 4}}=\frac{1}{2(2 \pi)^3 a^3}\int d^3 k \, e^{i {\bf k}{\bf {\cdot }} \Delta {\bf x}}(\Omega_{{\bf k}}^{-1})^{(4)}
\ee
can be split into two parts as follows:
\begin{subequations}
\be
I^{(4)}_{{\bf 4}}=I^{(4)}_{\mathrm {EM}}+I^{(4)}_{\mathrm {GR}},
\ee
where
\bea
I^{(4)}_{\mathrm {EM}}&=&\frac{K_{1}(m a \epsilon)}{576  \pi ^2 m a \epsilon}\left[-12 m^2 a \epsilon ^2 \Delta x^2
   \dot{a} q^2 A \dot{A}+6 \epsilon^2 q^2\dot{A}^2+q^2 A^2
   m^2 \Delta x^2(6 q^2 A^2 \Delta x^2-a^2 \epsilon^2
   R)\right] \nonumber \\
   & & +\frac{K_0(m a \epsilon)}{192 \pi
   ^2} \Delta x^2 \left[4  \frac{\dot{a}}{a}
   q^2 A \dot{A}+ 2 q^2\dot{A}^2+4 q^2A
   \ddot{A}+q^2 A^2 \left(m^2 \epsilon^2
   \dot{a}^2+2 (6 \xi -1)
   R\right)\right]\, ,
\\
I^{(4)}_{\mathrm {GR}}&=&\frac{K_0(m a \epsilon)}{5760  \pi^2}a^2 \epsilon^2 \left[6 (43-240 \xi )  \frac{\dot{a}^2 \ddot{a}}{a^3}- 11 m^2  \frac{\epsilon^2 \dot{a}^2 \ddot{a}}{a}- 7 m^2 \frac{\epsilon^2  \dot{a}^4}{a^2}+18 (3-20 \xi )   \frac{\ddot{a}^2}{a^2}+6 (7-60 \xi ) \frac{\dot{a}   a^{(3)}}{a^2}-6  \frac{a^{(4)}}{a}\right]\nonumber\\
&&+\frac{ K_1(m a \epsilon)}{23040 
   m \pi ^2} a \epsilon \left\{ \left[5  m^4 a^4 \epsilon^4+720
   (1-6 \xi )^2+16 m^2 a^2 \epsilon^2 (45
   \xi -8)\right] \frac{\dot{a}^4}{a^4}+48 \left(a^2 m^2
   \epsilon^2-18+90 \xi \right) \frac{\dot{a}
   a^{(3)}}{a^2} \right.  \nonumber \\
   &&\qquad \qquad \qquad \qquad  \left. +8
   \left[348+180 \xi  (36 \xi -17 )+
   m^2 a^2\epsilon^2 (1+90 \xi )\right] \frac{\dot{a}^2
   \ddot{a}}{a^3}+288  (5 \xi -1 )
   \frac{a^{(4)}}{a} \right. \nonumber\\
   &&\qquad \qquad \qquad \qquad \left. +36\left[12+a^2
   m^2 \epsilon^2+40 \xi  (18 \xi-5 )\right]
   \frac{\ddot{a}^2}{a^2}\right\} \, .
\eea
\end{subequations}

\section{$\sigma$, $\Delta ^{\frac{1}{2}}$}
\label{ap:geodesic-dist}

The defining equation for the geodesic interval $\sigma$ is 
\be
2 \sigma=g_{\mu \nu}\sigma^{;\mu}\sigma^{;\nu}.
\ee
For the space-time defined by the metric \eqref{FLRWmetric},  the above relation takes the form
\be 
2\sigma=-(\partial_t \sigma)^2+\frac{1}{a^2}(\partial_i \sigma)^2\equiv-(\partial_t \sigma)^2+\frac{1}{a^2}(\partial_\epsilon \sigma)^2. \label{eq:sigma-FLRW}
\ee
For small space-time separations, we propose the following ansatz
\be \label{eq:ansatz-sigma}
\sigma=\sum_{n=2}^{\infty}\sum_{i=0}^{n}c_{n,i}(t)(\Delta t)^i (\epsilon)^{n-i}
\ee
where the coefficients $c_{n,i}$ depend only on the time $t$ and satisfy the boundary conditions $c_{2,2}=-\frac{1}{2}$, $c_{2,1}=0$ and $c_{2,0}=\frac{a^2}{2}$. Although we are interested in the case with $\Delta t =0$, to compute a short-distance expansion for $\sigma $ we need to include all the above terms.

Inserting \eqref{eq:ansatz-sigma} into \eqref{eq:sigma-FLRW} and grouping terms with the same short-distance order $n$, we can iteratively obtain the higher order terms from the lowest orders. 
The first few non-vanishing coefficients are
\begin{align}
&c_{3,1}=-\frac{1}{2}a \dot a\, , \quad
c_{4,0}=\frac{1}{24}a^2 \dot a^2\, , \quad c_{4,2}=\frac{1}{6}a \ddot a\, ,\nonumber \\
&c_{5,1}=-\frac{1}{24}\big(a \dot a^3+a^2 \dot a \ddot a\big)\, , \quad
c_{5,3}=\frac{1}{24}\big(\dot a \ddot a -  a a^{(3)} \big)\, , \nonumber \\
&c_{6,0}=\frac{1}{720}\big(a^2 \dot a^4 + 3 \dot a^2 \ddot a\big)\, , \nonumber \\
&c_{6,2}=\frac{1}{720}\big(31 a \dot a^2 \ddot a+8 a^2 \ddot a^2 +9a^2 \dot a a^{(3)} \big)\, , \nonumber \\
&c_{6,4}=\frac{1}{360 a}\big(6 \dot a^2 \ddot a-7 a \ddot a^2-6a \dot a a^{(3)}+3 a^2 a^{(4)}\big)\, .
\end{align}
The defining equation for the Van Vleck-Morette determinant  $\Delta ^{\frac{1}{2}}$ is
\be
\nabla_{\mu} \nabla^{\mu} \sigma=N-2 \Delta^{-\frac{1}{2}} \Delta_{; \mu}^{\frac{1}{2}} \sigma^{; \mu}\,.
\ee
In flat FLRW coordinates and in terms of $\epsilon$, the above equation reads 
\be
-\partial_{t}^{2} \sigma-(N-1)\frac{ \dot{a}}{a} \partial_{t} \sigma+(N-2)\frac{\partial_{\epsilon}\sigma}{a^2\epsilon}+\frac{\partial_{\epsilon}^2 \sigma}{a^2}=N+\frac{2}{\Delta^{\frac{1}{2}}}\Big(\partial_{t} \Delta^{\frac{1}{2}} \partial_{t} \sigma-\frac{\partial_{\epsilon} \Delta^{\frac{1}{2}} \partial_{\epsilon} \sigma}{a^{2}}\Big)\,.
\ee
As for the geodesic distance, we propose the following short-distance expansion for $\Delta^{\frac{1}{2}}$
\be \label{eq:ansatz-delta12}
\Delta^{\frac{1}{2}}=1+\sum_{n=2}^{\infty}\sum_{i=0}^{n}f_{n,i}(t) (\Delta t)^{i} (\epsilon)^{n-i}\, ,
\ee
where again the coefficients $f_{n,i}$ depend only on $t$.
Inserting the ansatz \eqref{eq:ansatz-delta12} together with \eqref{eq:ansatz-sigma} into the defining equation, we can iteratively find the coefficients of the expansion. The second order coefficients are
\bea
f_{2,0}=\frac{1}{2}(a \ddot a+(N-2)\dot a^2)\, , \quad f_{2,1}=0\, , \quad f_{2,2}=-\frac{(N-1)\ddot a}{12 a}\,. 
\eea
As expected, our expansion coincides with the short-distance covariant expansion of the Van Vleck-Morette determinant, given, for example, in \cite{Decanini:2005eg}.

\begin{acknowledgments}
We thank J.~Navarro-Salas and P.R.~Anderson for many enlightening discussions.
We also thank the anonymous referee for their careful reading of this manuscript and helpful comments which have improved the paper, particularly in the interpretation of our results in \eqref{eq:WN3}.
Most of the computations in this paper have been done with the aid of the xAct {\tt {MATHEMATICA}} package~\cite{xAct-1}.
The work of S.P.~is supported by the Leverhulme Trust, Grant No.~RPG-2021-299, and was previously supported by the Ministerio de Ciencia, Innovaci\'on y Universidades, Ph.D. fellowship, Grant No. FPU16/05287. 
	The work of E.W.~is supported by the Lancaster-Manchester-Sheffield Consortium for Fundamental Physics under STFC grant ST/T001038/1. 
	This research has also received funding from the European Union's Horizon 2020 research and innovation program under the H2020-MSCA-RISE-2017 Grant No.~FunFiCO-777740.
 S.P. thanks the Gravitation and Cosmology Research Group (CRAG) for their hospitality during her visit to the University of Sheffield, where part of this work was carried out. 
	No data were created or analysed in this study.
\end{acknowledgments}

\bibliography{refs}

\end{document}